\documentclass[final]{aipproc}
\layoutstyle{6x9}
\begin{document}
\title{\textbf{Gravitational energy as dark energy:\\ cosmic structure and
apparent acceleration}\footnote{Based on a presentation at the {\em
International Conference on Two Cosmological Models}, Universidad
Iberoamericana, Mexico City, 17-19 November, 2010; to appear in the
Proceedings, ed.~J.~Auping.}}
\classification{98.80.-k 98.80.Es 95.36.+x 98.80.Jk}
\keywords{dark energy, theoretical cosmology, observational cosmology}
\author{David L. Wiltshire}{
address={Department of Physics \& Astronomy, University of Canterbury, Private Bag 4800,\\ Christchurch 8140, New Zealand}
}

\begin{abstract}
Below scales of about $100\,h^{-1}\,$Mpc our universe displays a complex
inhomogeneous structure dominated by voids, with clusters of galaxies in
sheets and filaments. The coincidence that cosmic expansion appears to start
accelerating at the epoch when such structures form has prompted a number
of researchers to question whether dark energy is a signature of a failure
of the standard cosmology to properly account, on average, for the
distribution of matter we observe. Here I discuss the timescape scenario,
in which cosmic acceleration is understood as an apparent
effect, due to gravitational energy gradients that grow when spatial
curvature gradients become significant with the nonlinear growth of cosmic
structure. I discuss conceptual issues related to the averaging problem,
and their impact on the calibration of local geometry to the
solutions of the volume--average evolution equations corrected by
backreaction, and the question of nonbaryonic dark matter in the timescape
framework. I further discuss recent work on defining observational
tests for average geometric quantities which can distinguish the timescape
model from a cosmological constant or other models of dark energy.
\end{abstract}
\maketitle
\font\fiverm=cmr5\font\sevenrm=cmr7\font\sevenit=cmmi7
\def\sbf{\footnotesize\bf}
\def\ns#1{_{\hbox{\sevenrm #1}}}
\def\PRL#1{{\em Phys.\ Rev.\ Lett.}\ {\bf#1}}
\def\PL#1{{\em Phys.\ Lett.}\ {\bf#1}}
\def\PR#1{{\em Phys.\ Rev.}\ {\bf#1}}
\def\ApJ#1{{\em Astrophys.\ J.}\ {\bf#1}}
\def\AsJ#1{{\em Astron.\ J.}\ {\bf#1}}
\def\AaA#1{{\em Astron.\ Astrophys.}\ {\bf#1}}
\def\MNRAS#1{{\em Mon.\ Not.\ R.\ Astr.\ Soc.}\ {\bf#1}}
\def\CQG#1{{\em Class.\ Quantum Grav.}\ {\bf#1}}
\def\GRG#1{{\em Gen.\ Relativ.\ Grav.}\ {\bf#1}}
\def\IJMP#1{{\em Int.\ J.\ Mod.\ Phys.}\ {\bf #1}}
\def\JCAP#1{{\em JCAP}\ {#1}}
\def\beq{\begin{equation}} \def\eeq{\end{equation}}
\def\bea{\begin{eqnarray}} \def\eea{\end{eqnarray}}
\def\Z#1{_{\lower2pt\hbox{$\scriptstyle#1$}}} \def\w#1{\,\hbox{#1}}
\def\X#1{_{\lower2pt\hbox{$\scriptscriptstyle#1$}}}
\font\sevenrm=cmr7 \def\ns#1{_{\hbox{\sevenrm #1}}} \def\dOM{\dd\Omega^2}
\def\Ns#1{\Z{\hbox{\sevenrm #1}}} \def\ave#1{\langle{#1}\rangle}
\def\lsim{\mathop{\hbox{${\lower3.8pt\hbox{$<$}}\atop{\raise0.2pt\hbox{$\sim$}}
$}}} \def\kmsMpc{\w{km}\;\w{sec}^{-1}\w{Mpc}^{-1}} \def\bn{\bar n}
\def\dd{{\rm d}} \def\ds{\dd s} \def\etal{{\em et al}.}
\def\al{\alpha}\def\be{\beta}\def\ga{\gamma}\def\de{\delta}\def\ep{\epsilon}
\def\et{\eta}\def\th{\theta}\def\ph{\phi}\def\rh{\rho}\def\si{\sigma}
\def\gsim{\mathop{\hbox{${\lower3.8pt\hbox{$>$}}\atop{\raise0.2pt\hbox{$
\sim$}}$}}} \def\ta{\tau} \def\ac{a}
\def\frn#1#2{{\textstyle{#1\over#2}}} \def\Deriv#1#2#3{{#1#3\over#1#2}}
\def\Der#1#2{{#1\hphantom{#2}\over#1#2}} \def\pt{\partial} \def\ab{{\bar a}}
\def\tw{\ta}\def\gb{\bar\ga} \def\BB{{\cal B}} \def\CC{{\cal C}}
\def\av{{a\ns{v}\hskip-2pt}} \def\aw{{a\ns{w}\hskip-2.4pt}}\def\Vav{{\cal V}}
\def\DD{{\cal D}}\def\gd{{{}^3\!g}}\def\half{\frn12}\def\Rav{\ave{\cal R}}
\def\QQ{{\cal Q}}\def\dsp{\displaystyle} \def\rw{r\ns w}
\def\mean#1{{\vphantom{\tilde#1}\bar#1}} \def\bx{{\mathbf x}}
\def\bH{\mean H}\def\Hb{\bH\Z{\!0}}\def\bq{\mean q}
\def\gb{\mean\ga}\def\gc{\gb\Z0} \def\OMMn{\OM\Z{M0}}
\def\rhb{\mean\rh}\def\OM{\mean\Omega}\def\etb{\mean\eta}
\def\fw{{f\ns w}}\def\fv{{f\ns v}} \def\goesas{\mathop{\sim}\limits}
\def\fvn{f\ns{v0}} \def\fvf{\left(1-\fv\right)} \def\Hh{H}
\def\OMM{\OM\Z M}\def\OmMn{\Omega\Z{M0}} \def\ts{t}\def\tb{\ts'} \def\Hm{H\Z0}
\def\Fi{\hbox{\footnotesize\it fi}}\def\etw{\eta\ns w} \def\etv{\eta\ns v}
\def\fvi{{f\ns{vi}}} \def\fwi{{f\ns{wi}}}
\def\Hx{H_{\lower1pt\hbox{$\scriptstyle0$}}}
\def\LCDM{$\Lambda$CDM} \def\OmLn{\Omega\Z{\Lambda0}}\def\Omkn{\Omega\Z{k0}}
\def\Dtc{\mathop{\hbox{$\Der\dd\tw$}}}
\def\Dfb{D\Ns{TS}} \def\Dlcdm{D\Ns{$\scriptstyle\Lambda$CDM}}
\def\dL{d\Z L} \def\dA{d\Z A} \def\name{timescape}
\def\OmBn{\Omega\Z{B0}} \def\OmCn{\Omega\Z{C0}}
\def\CT{Cooperstock--Tieu}\def\Msun{M_\odot}
\section{Introduction}

This conference is titled a ``Conference on Two Cosmological Models''
but I think that what have been presented are certainly more than
two cosmological models. From the point of view of proponents of the
standard cosmology, the conference might seem to be dealing with
too many cosmological models. The fact that there are a lot of ideas on
the table is natural at any point in the history of science when
observations present a fundamental crisis. We have reached such a
point, given that our current standard cosmology only works by
invoking unknown sources of ``dark energy'' and ``dark matter'', which
supposedly make up most of the stuff in the universe.

All of the models presented, including the standard \LCDM\ cosmology, could
be said to be relativistic in the sense that they obey either Einstein's
equations or some extension of Einstein gravity with
a geometric diffeomorphism invariant action. What is at issue, however,
is the manner in which different models seek to explain the observed
large scale structure and motion of objects in the universe. Do we add
new fields or modifications to the gravitational action, whose only influence
is on cosmological scales, or do we
seek to find deeper answers in the principles of general relativity?

These are questions that Einstein struggled with when he first applied
general relativity to cosmology \cite{esu}. He thought of spacetime as
being a relational structure, and therefore the introduction of a
cosmological constant -- a vacuum energy in the fabric of space which
made no direct connection to inertial properties of matter -- was not
a step he took lightly. I will take the viewpoint that rather than
adding further epicycles to the gravitational action, the cosmological
observations which we currently interpret in terms of dark energy are
inviting us to think more deeply about the foundations of general
relativity. There are questions in general relativity -- relating
to coarse--graining, averaging and the definition of energy in such
contexts -- which have never been fully resolved. These are the questions
which I believe are of relevance to cosmology.

In this paper I will review the conceptual basis \cite{clocks,equiv,fqxi} and
observational tests \cite{obs} of a cosmology model \cite{clocks,sol}, which
represents a new approach to understanding the phenomenology of dark energy
as a consequence of the effect of the growth of inhomogeneous structures.
The basic idea, outlined in a nontechnical manner in ref.\ \cite{dark07},
is that as inhomogeneities grow one must consider not only their backreaction
on average cosmic evolution, but also the variance in the geometry as it
affects the calibration of clocks and rulers of ideal observers. Dark energy is
then effectively realised as a misidentification of gravitational energy
gradients.

Although the standard Lambda Cold Dark Matter (\LCDM) model provides a good
fit to many tests, there are tensions between some tests, and also a number
of puzzles and anomalies. Furthermore, at the present epoch the observed
universe is only statistically homogeneous once one samples on scales
of 150--300 Mpc. Below such scales it displays a web--like structure,
dominated in volume by voids. Some 40\%--50\% of the volume of the present
epoch universe is in voids with $\de\rho/\rho\goesas-1$ on scales of
30$h^{-1}$ Mpc \cite{HV}, where $h$ is the dimensionless parameter related to
the Hubble constant by $H\Z0=100h\kmsMpc$. Once one also accounts for
numerous minivoids, and perhaps also a few larger voids, then it appears that
the present epoch universe is void-dominated. Clusters of galaxies are spread
in sheets that surround these voids, and in thin filaments that thread them.

A number of different approaches have been taken to study inhomogeneous
cosmologies. One large area of research is that of exact solutions of
Einstein's equations (see, e.g., ref.~\cite{BKHC}), and of the
Lema\^{\i}tre--Tolman--Bondi \cite{LTB} (LTB) dust solution in particular.
While one may mimic any luminosity distance relation with LTB models,
generally the inhomogeneities required to match type Ia supernovae (SneIa)
data are much larger than the typical scales of voids
described above. Furthermore, one must assume the unlikely symmetry of a
spherically symmetric universe about our point, which violates the Copernican
principle. It is my view that while the LTB solutions are interesting toy
models, one should retain the Copernican principle in a statistical sense,
and one should seriously try to model the universe with those scales of
inhomogeneity that we actually observe.

One particular consequence of a matter distribution that is only
statistically homogeneous, rather than exactly homogeneous, is that when
the Einstein equations are averaged they do not evolve as a smooth
Friedmann--Lema\^{\i}tre--Robertson--Walker (FLRW) geometry. Instead
the Friedmann equations are supplemented by additional backreaction
terms\footnote{For a general review of averaging and backreaction see, e.g.,
ref.\ \cite{vdH}.} \cite{buch00}.
Whether or not one can fully explain the expansion history of the universe
as a consequence of the growth of inhomogeneities and backreaction, without
a fluid--like dark energy, is the subject of ongoing debate \cite{buch08}.

A typical line of reasoning against backreaction is that of a plausibility
argument \cite{Peebles}: if we {\em assume}
a FLRW geometry with small perturbations, and estimate the magnitude of
the perturbations from the typical rotational and peculiar velocities of
galaxies, then the corrections of inhomogeneities are consistently small. This
would be a powerful argument, were it not for the fact that at the present
epoch galaxies are not homogeneously distributed. The Hubble Deep Field
reveals that galaxy clusters were close to being homogeneous distributed at
early epochs, but following the growth voids at redshifts $z\lsim1$
that is no longer the case today. Therefore galaxies cannot be consistently
treated as randomly distributed gas particles on the 30$h^{-1}$ Mpc scales
\cite{HV} that dominate present cosmic structure below the scale of
statistical homogeneity.

Over the past few years I have developed a new physical interpretation of
cosmological solutions within the Buchert averaging scheme
\cite{clocks,equiv,sol}. I start
by noting that in the presence of strong spatial curvature
gradients, not only should the average evolution equations be replaced
by equations with terms involving backreaction, but the physical
interpretation of average quantities must also account for the differences
between the local geometry and the average geometry.
In other words, geometric variance can be just as important as
geometric averaging when it comes to the physical interpretation of the
expansion history of the universe.

I proceed from the fact that structure formation provides a natural division
of scales in the observed universe. As observers in galaxies, we and the
objects we observe in other galaxies are necessarily in bound structures,
which formed from density perturbations that were greater than critical
density. If we consider the evidence of the large scale structure
surveys on the other hand, then the average location by volume in the
present epoch universe is in a void, which is negatively curved.
We can expect systematic
differences in spatial curvature between the average mass environment, in
bound structures, and the volume-average environment, in voids.

Spatial curvature gradients will in general give rise to gravitational
energy gradients, and herein lie the issues which I believe are key
to understanding the phenomenon of dark energy. The definition of
gravitational energy in general relativity is notoriously subtle.
This is due to the equivalence principle, which
means that we can always get rid of gravity near a
point. As a consequence, the energy, momentum and angular momentum associated
with the gravitational field, which have macroscopic effects on the relative
calibrations of the clocks and rulers of observers, cannot be described
by local quantities encoded in a fluid-like energy-momentum tensor.
Instead they are at best {\em quasi-local} \cite{quasi_rev}. There is
no general agreement on how to deal with quasi-local gravitational
energy. It is my view that since the issue has its origin in the
equivalence principle, we must return to first principles and reconsider
the equivalence principle in the context of cosmological averages.

\section{The cosmological equivalence principle}

In laying the foundations of general relativity, Einstein sought
to refine our physical understanding of that most central physical concept:
{\em inertia}. As he stated: ``In a consistent theory of relativity
there can be be no inertia relatively to `space', but only an inertia of
masses relatively to one another'' \cite{esu}. This is the general
philosophy that underlies Mach's principle, which strongly guided Einstein.
However, the refinement of the understanding of inertia that Einstein
left us with in relation to gravity, the Strong Equivalence Principle
(SEP), only goes part-way in addressing Mach's principle.

Mach's principle may be stated
\cite{Bondi,BKL}: {\em ``Local inertial frames (LIFs) are determined
through the distributions of energy and momentum in the universe by some
weighted average of the apparent motions''}. The SEP says nothing about
the average effect of gravity, and therefore nothing about the suitable
``weighted average of the apparent motions'' of the matter in the universe.
Since gravity for ordinary matter fields obeying the strong energy condition
is universally attractive, the spacetime geometry of a universe containing
matter is not stable, but is necessarily dynamically evolving. Therefore,
accounting for the average effect of matter to address Mach's principle
means that any relevant frame in cosmological averages is one in which time
symmetries of the Lorentz group in LIFs are removed.

My proposal for applying the equivalence principle on cosmological scales
is to deal with the average effects of the evolving density by extending the
SEP to larger regional frames while removing the time translation and boost
symmetries of the LIF to define a {\em Cosmological Equivalence Principle}
as follows \cite{equiv}:

{\em At any event, always and everywhere, it is possible to choose a suitably
defined spacetime neighbourhood, the cosmological inertial frame (CIF), in
which average motions (timelike and null) can be described by geodesics in a
geometry that is Minkowski up to some time-dependent conformal
transformation},
\beq \ds^2\Ns{CIF}=
a^2(\et)\left[-\dd\et^2+\dd r^2+r^2(\dd\th^2+\sin^2\th\,\dd\ph^2)\right].
\label{cif}\eeq

Since the average geometry is a time--dependent conformal scaling of Minkowski
space, the CEP reduces to the standard SEP if $a(\et)$ is constant, or
alternatively over very short time intervals during which the time variation
of $a(\et)$ can be neglected. The relation to cosmological averages is
understood by the fact that (\ref{cif}) is the spatially flat FLRW
metric. In the standard cosmology this is taken to be the geometry of
the whole universe. Here, however, the whole universe is inhomogeneous
while its geometry is restricted by the requirement that it is possible
to always choose (\ref{cif}) as a regional average. This would
rule out geometries with global anisotropies, such as Bianchi models,
while hopefully leaving enough room to describe an inhomogeneous but
statistically homogeneous universe like the one we observe.

To understand why an average geometry (\ref{cif}) is a relevant average
reference geometry for the relative calibration of rulers and clocks in the
absence of global Killing vectors, let us construct what I will call the
{\em semi-tethered lattice} by the following means. Take a lattice of
observers in Minkowski space, initially moving isotropically away from each
nearest neighbour at uniform initial velocities. The lattice of observers
are chosen to be equidistant along mutual oriented $\hat x$, $\hat y$ and
$\hat z$ axes. Now suppose that the observers are each connected to
six others by tethers of negligible mass and identical tension along the
mutually oriented spatial
axes. The tethers are not fixed but unwind
freely from spools on which an arbitrarily long supply of tether is wound.
The tethers initially
unreel at the same uniform rate, representing a ``recession velocity''.
Each observer carries synchronised clocks, and at a prearranged local proper
time all observers apply brakes to each spool,
the braking mechanisms having
been pre-programmed to deliver the same impulse as a function of local
time.

The semi-tethered lattice experiment is directly analogous to
the decelerating volume expansion of (\ref{cif}) due to some average
homogeneous matter density, because it maintains the homogeneity and
isotropy of space over a region as large as the lattice. Work is done in
applying the brakes, and energy can be extracted from this -- just as
kinetic energy of expansion of the universe is converted to other
forms by gravitational collapse. Since brakes are applied in unison, however,
there is {\em no net force on any observer in the lattice}, justifying the
{\em inertial frame} interpretation, even though each observer has a
nonzero 4-acceleration with respect to the global Minkowski frame.
The braking function may have an arbitrary time profile; provided it
is applied uniformly at every lattice site the clocks will remain
synchronous in the comoving sense, as all observers
have undergone the same relative deceleration.

Whereas the Strong Equivalence Principle leads us to define local inertial
frames, related to each other by local Lorentz transformations acting at
a point, the Cosmological Equivalence Principle refers to a {\em collective}
symmetry of the background. In defining the averaging region of the CIF
we are isolating just that part of the volume expansion which is regionally
homogeneous and isotropic, and which is determined by the regionally
homogeneous part of the background density.

\begin{figure}[htb]
\centerline{\ {\sbf(a)}\hskip-20pt
\includegraphics[width=2.5in]{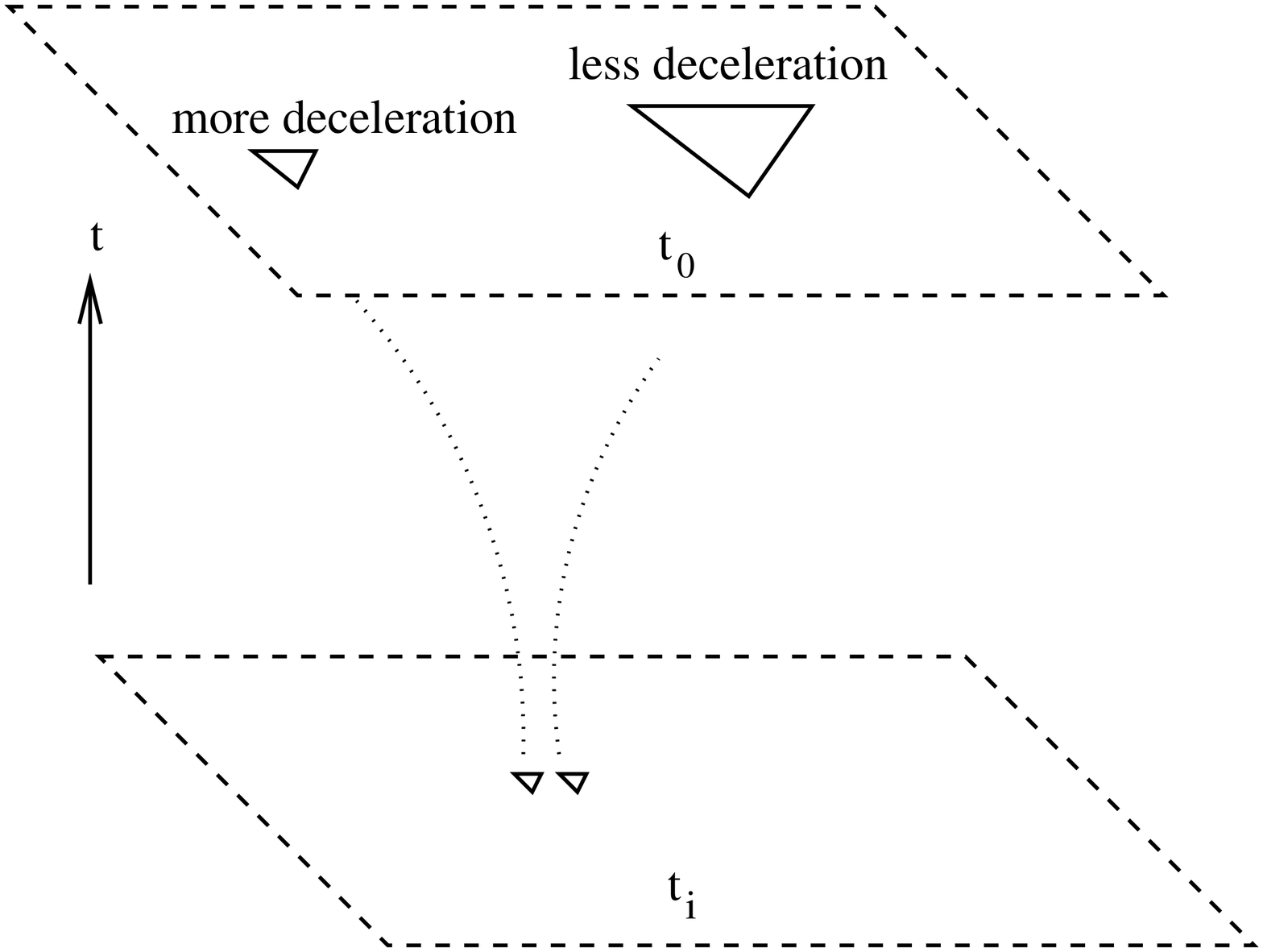}
\ {\sbf(b)}\hskip-20pt
\includegraphics[width=2.5in]{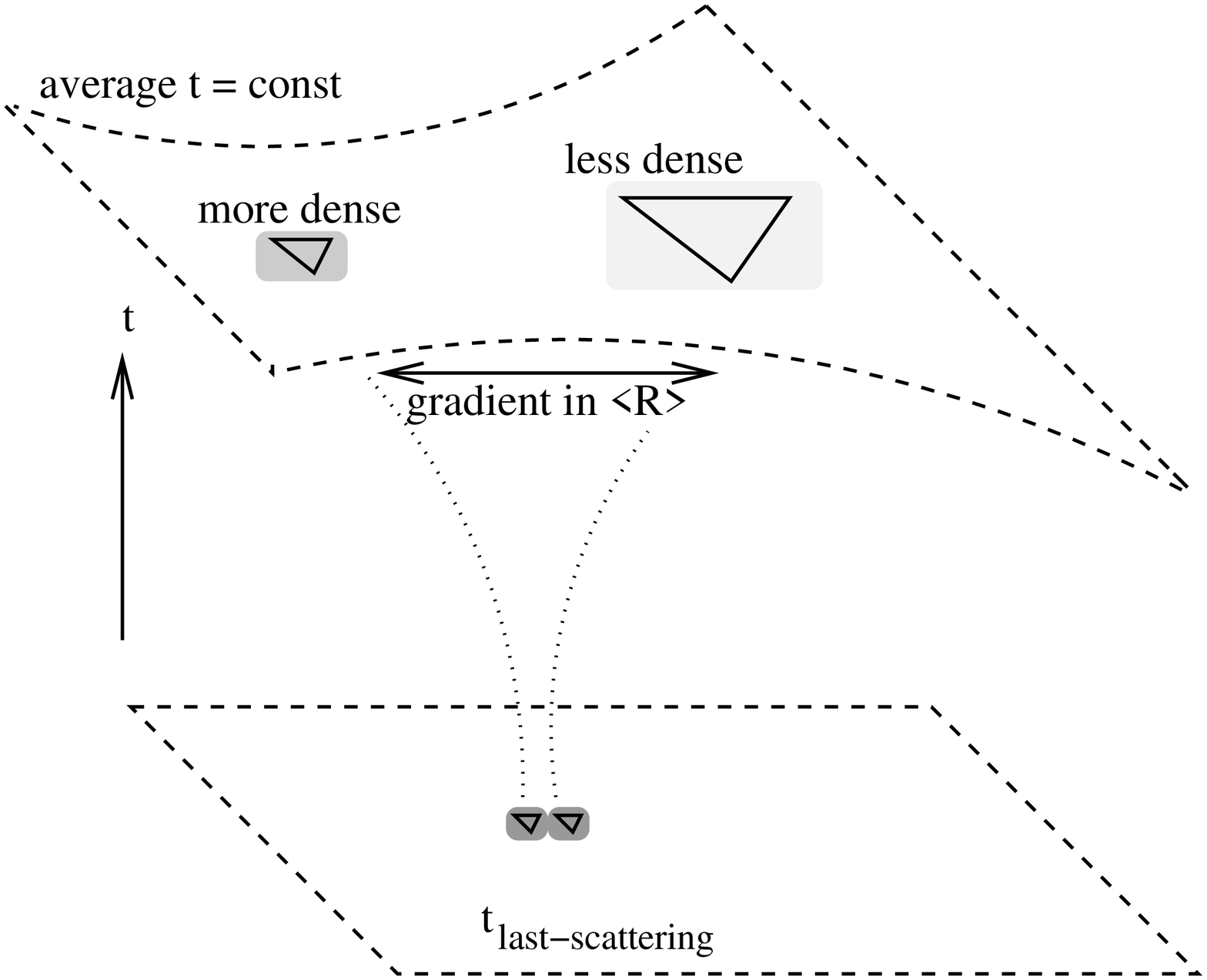}}
\caption{%
Two equivalent situations: {\bf(a)} in Minkowski space observers
in separate semi--tethered lattices, initially expanding at the same
rate, apply brakes homogeneously and isotropically within their respective
regions but at different rates;
{\bf(b)} in the universe which is close to
homogeneous and isotropic at last-scattering comoving observers in separated
regions initially move away from each other isotropically, but experience
different locally homogeneous isotropic decelerations as local density
contrasts grow. In both cases there is a relative deceleration
of the observer congruences and those in the region which has decelerated
more will age less.}
\label{fig_equiv}
\end{figure}

Let us now consider two sets of disjoint semi-tethered lattices,
with identical initial local expansion velocities, in a background
static Minkowski space. (See Fig.~\ref{fig_equiv}(a).)
Observers in the first congruence apply brakes in unison to decelerate
homogeneously and isotropically at one rate. Observers in the second
congruence do so similarly, but at a different rate.
Suppose that when transformed to a global Minkowski frame,
with time $t$,
that at each time step the magnitudes of the 4--decelerations satisfy
$\al\Z1(t)>\al\Z2(t)$ for the respective congruences. By special relativity,
since members of the first congruence decelerate more than those of the
second congruence, at any time $t$ their proper times satisfy $\ta\Z1<\ta\Z2$.
The members of the first congruence age less quickly than members of the
second congruence.

By the CEP, the case of volume expansion of
two disjoint regions of different average density in the actual universe
is entirely analogous. The equivalence of the circumstance rests on the
fact that the expansion of the universe was extremely uniform at the time
of last scattering, by the evidence of the CMB. At that epoch all regions
had almost the {\em same} density -- with tiny fluctuations -- and the same
uniform Hubble flow. At late epochs, suppose that in the frame of any
average cosmological observer there are expanding regions of {\em different}
density which have decelerated by different amounts by a given time, $t$,
according to that observer. Then by the CEP the local proper time of the
comoving observers in the denser region, which has decelerated more,
will be less than that of the equivalent observers in the less
dense region which has decelerated less. (See Fig.~\ref{fig_equiv}(b).)
Consequently the {\em proper time of the observers in the more
dense CIF will be less than that of those in the less dense CIF}, by
equivalence of the two situations.

The fact that a global Minkowski observer does not exist in the second
case does not invalidate the argument. The global Minkowski time is
just a coordinate label. In the cosmological case the only restriction
is that {\em the expansion of both average congruences
must remain homogeneous and isotropic in local regions of different average
density} in the global average $t=$const slice. Provided we can patch the
regional frames together suitably, then if regions
in such a slice {\em are still expanding} and have a significant density
contrast we can expect a significant clock rate variance.

This equivalence directly establishes the idea of a {\em gravitational
energy cost for a spatial curvature gradient}, since the existence
of expanding regions of different density within an average $t=$const
slice implies a gradient in the average Ricci scalar curvature, $\Rav$,
on one hand, while the fact that the local proper time varies
on account of the relative deceleration implies a gradient in gravitational
energy on the other.

In the actual universe, the question is: can the effect described above
be significant enough to give a significant variation in the clocks of
ideal isotropic observers (those who see an isotropic mean CMB) in
regions of different density, who experience a relative deceleration
of their regional volume expansions? Since we
are dealing with weak fields the relative deceleration of the background
is small. Nonetheless even if the relative deceleration is typically of
order $10^{-10}$ms$^{-2}$, cumulatively over the age of the universe
it leads to significant clock rate variances \cite{equiv}, of the
order of 38\%. Such a large effect is counterintuitive,
as we are used to only considering time dilations due to relative
accelerations within the static potentials of isolated systems.
Essentially, we are dealing with a different physical effect concerning
the relative synchronization of clocks in the absence of
global Killing vectors. A small instantaneous relative deceleration
can lead to cumulatively large differences, given one has the lifetime
of the universe to play with.
As a consequence the age of the universe itself becomes position--dependent.
Since we and all the objects we observe are necessarily in regions of
greater than critical density, on account of structure formation we have a
mass--biased view of the universe and cannot directly observe such
variations.

In the standard ADM formalism one assumes the existence of a global
rest frame comoving with the dust, and one makes a ($3+1$)--split of the
Einstein equations from the point of view of fundamental observers who
may be either comoving or tilted with respect to the dust. If the only
symmetries that are allowed are diffeomorphisms of the global metric on one
hand, and local Lorentz transformations corresponding to rotations and boosts
on the other, then realistically there is no room within such an ADM formalism
for clock rate variations of the order of magnitude dealt with in the
timescape scenario \cite{R10}. However, such criticism overlooks the very real
possibility the rest frame of dust is not globally defined, and furthermore it
overlooks the crucial idea of regional averages introduced by the CEP. In
proposing to separate the collective degree of freedom of the quasi-local
regional volume expansion from other gravitational degrees of freedom, I am
suggesting that we must consider average regional symmetries as a
completely new ingredient in addition to the global diffeomorphisms and
local Lorentz transformations with which we are familiar. This is a potential
route to dealing with the unsolved problems of coarse-graining in
general relativity.

At the epoch of last scattering dust may certainly be assumed to be atomic.
However, once structures form geodesics cross and at the present epoch dust
must be coarse-grained on at least the scale of galaxies in cosmological
modelling. Thus the issue of the coarse-graining of dust is not merely a matter
of choice, but of physical necessity if one is to consistently think about the
interpretation of the Buchert formalism\footnote{One can apply the Buchert
formalism in a different manner -- for example, to the problem of prescribed
dust in exact solutions such as the LTB model \cite{BA}--\cite{MM} on globally
well-defined spacelike hypersurfaces, where one specifically avoids solutions
which develop vorticity or singularities. However, it is my view that to deal
with the actual inhomogeneities of the observed universe then the average
evolution of the Einstein equations should be regarded as a statistical
description, and I approach the Buchert formalism in this sense.}. Of course, a
detailed mathematical framework\footnote{For one approach to coarse-graining,
as opposed to averaging, see ref.\ \cite{Kor09}.} for this still
remains to be given in the \name\ scenario. However, it is my view that
mathematics is best guided by physical intuition rather than the reverse,
and consequently my work to date has proceeded from making a phenomenological
ansatz consistent with the CEP, to see whether the idea stands a chance
of working.

\section{The timescape model}
I proceed from an ansatz that the variance in gravitational energy is
correlated with the average spatial curvature in such a way as to implicitly
solve the Sandage--de Vaucouleurs paradox that a statistically quiet,
broadly isotropic, Hubble flow is observed deep below the scale of
statistical homogeneity. In particular, galaxy peculiar velocities
have a small magnitude with respect to a local regional volume expansion.
Expanding regions of different densities are patched together so that the
regionally measured expansion remains uniform. Such regional expansion
refers to the variation of the regional proper length, $\ell_r=\Vav^{1/3}$,
with respect to proper time of isotropic observers.
Although voids open up faster, so that their proper volume increases more
quickly, on account of gravitational energy gradients the local clocks will
also tick faster in a compensating manner.

In order to deal with dust evolution from the surface of last scattering up to
the present epoch, I assume that dust can be coarse--grained at the
$100h^{-1}$Mpc scale of statistical homogeneity over which mass flows can be
neglected. The manner in which I interpret the Buchert formalism is therefore
different to that adopted by Buchert \cite{buch00}, who does not define the
scale of coarse-graining of the dust explicitly. Details of the fitting of
local observables to average quantities for solutions to the Buchert
equations\footnote{The model of Wiegand and Buchert \cite{WB}, briefly
described by Buchert in the present volume \cite{buch10}, has similarities to
the present model but also differs from it in certain key aspects. In
particular, (i) the observational interpretation of the Buchert averages is
different; (ii) the walls and voids are taken to have internal backreaction in
the case or refs.\ \cite {WB,buch10} but not here; and (iii) the interpretation
of the initial wall and void fractions at last scattering is different.
In respect of the last point, since walls and voids do not exist at the
surface of last scattering, I take the view that the vast bulk of the
present horizon volume that averages to critical density gives $\fwi\simeq1$,
while $\fvi=1-\fwi$ is the small positive fraction of the present horizon
volume that consists of uncompensated underdense regions at last scattering
surface.} are described in detail in refs.\ \cite{clocks,sol}. Negatively
curved voids, and spatially flat expanding wall regions within which galaxy
clusters are located, are combined in a Buchert average
\beq\fv(t)+\fw(t)=1,\eeq
where $\fw(t)=\fwi\aw^3/\ab^3$ is the {\em wall volume fraction} and
$\fv(t)=\fvi\av^3/\ab^3$ is the {\em void volume fraction},
$\Vav=\Vav\ns i\ab^3$ being the present horizon volume, and $\fwi$, $\fvi$ and
$\Vav\ns i$ initial values at last scattering. The time parameter, $t$,
is the volume--average time parameter of the Buchert formalism, but does not
coincide with that of local measurements in galaxies. In trying to fit a
FLRW solution to the universe we attempt to
match our local spatially flat wall geometry
\beq\ds^2\Z{\Fi}=-\dd\tw^2+\aw^2(\tw)\left[\dd\etw^2+
\etw^2\dOM\right]\,.
\label{wgeom}\eeq
to the whole universe, when in reality the calibration of rulers and clocks of
ideal isotropic observers vary with gradients in spatial curvature and
gravitational energy.
By conformally matching radial null geodesics with those of the Buchert
average solutions, the geometry (\ref{wgeom}) may be extended to cosmological
scales as the dressed geometry
\beq
\ds^2=-\dd\tw^2+\ac^2(\tw)\left[\dd\etb^2+\rw^2(\etb,\tw)\,\dOM\right]
\label{dgeom}\eeq
where $a=\gb^{-1}\ab$, $\gb=\Deriv\dd\tw\ts$ is the relative lapse
function\footnote{This is a phenomenological function rather than the
lapse function prescribed by the ADM formalism.} between wall clocks and
volume--average ones, $\dd\etb=\dd t/\ab=\dd\tw/ \ac$, and
$\rw=\gb\fvf^{1/3}\fwi^{-1/3}\etw(\etb,\tw)$, where $\etw$ is given by
integrating $\dd\etw=\fwi^{1/3}\dd\etb/[\gb\fvf^{1/3}]$ along null geodesics.

In addition to the bare cosmological parameters which describe the Buchert
equations, one obtains dressed parameters relative to the geometry
(\ref{dgeom}). For example, the dressed matter density parameter is
$\Omega\Z M=\gb^3\OMM$, where $\OMM=8\pi G\rhb\Z{M0}\ab\Z0^3/(3\bH^2\ab^3)$
is the bare matter density parameter. The dressed parameters take numerical
values close to the ones inferred in standard FLRW models.

\subsection{Apparent acceleration and Hubble flow variance}

The gradient in gravitational energy and cumulative differences of clock
rates between wall observers and volume average ones has important
physical consequences. Using the exact solution
obtained in ref.\ \cite{sol}, one finds that a volume average observer
would infer an effective deceleration parameter $\bq=-\ddot\ab/(\bH^2\ab)=
2\fvf^2/(2+\fv)^2$, which is always positive since there is no global
acceleration. However, a wall observer infers a dressed deceleration
parameter
\beq
q={-1\over H^2 a}{\dd^2 a\over\dd\tw^2}=
{-\fvf(8\fv^3+39\fv^2-12\fv-8)\over\left(4+\fv+4\fv^2\right)^2}\,,
\label{qtrack}\eeq
where the dressed Hubble parameter is given by
\beq\Hh=\ac^{-1}\Dtc\ac=\gb\bH-\dot\gb\,=\gb\bH-\gb^{-1}\Dtc\gb\,.
\label{42}\eeq
At early times when $\fv\to0$ the dressed
and bare deceleration parameter both take the Einstein--de Sitter value
$q\simeq\bq\simeq\half$. However, unlike the bare parameter which
monotonically decreases to zero, the dressed parameter becomes negative
when $\fv\simeq0.59$ and $\bq\to0^-$ at late times. For the best-fit
parameters\footnote{Here I will simply adopt the parameters found in ref.\
\cite{LNW} from a fit to the Riess07 gold dataset \cite{Riess07}. A more recent
analysis \cite{SW} shows that the best-fit parameters are sensitive to the
method of supernova data reduction, and unknown systematic issues remain to
be resolved. The parameters determined from the Riess07 dataset are in the
mid--range of those determined by MLCS methods from larger datasets \cite{SW}.}
the apparent acceleration begins at a redshift $z\simeq0.9$.

Cosmic acceleration is thus revealed as an apparent effect which arises
due to the cumulative clock rate variance of wall observers relative to
volume--average observers. It becomes significant only when the voids
begin to dominate the universe by volume. Since the epoch of onset of
apparent acceleration is directly related to the void fraction, $\fv$, this
solves one cosmic coincidence problem.

In addition to apparent cosmic acceleration, a second important apparent
effect will arise if one considers scales below that of statistical
homogeneity. By any one set of clocks it will appear that voids expand
faster than wall regions. Thus a wall observer will see galaxies on the
far side of a dominant void of diameter $30h^{-1}$ Mpc recede at a
rate greater than the dressed global average $\Hm$, while galaxies within
an ideal wall will recede at a rate less than $\Hm$. Since the bare Hubble
parameter $\bH$ provides a measure of the uniform quasi-local flow, it
must also be the ``local value'' within an ideal wall at any epoch; i.e., eq.\
(\ref{42}) gives a measure of the variance in the apparent Hubble flow.
The best-fit parameters \cite{LNW} give a dressed Hubble
constant $\Hm=61.7^{+1.2}_{-1.1}\kmsMpc$, and a bare Hubble constant
$\Hb=48.2^{+2.0}_{-2.4}\kmsMpc$. The present epoch variance is 17--22\%.

Since voids dominate the universe by volume at the present epoch, any
observer in a galaxy in a typical wall region will measure locally higher
values of the Hubble constant, with peak values of order $72\kmsMpc$ at the
$30h^{-1}$ Mpc scale of the dominant voids. Over larger distances, as the
line of sight intersects more walls as well as voids, a radial spherically
symmetric average will give an average Hubble constant whose value decreases
from the maximum at the $30h^{-1}$ Mpc scale to the dressed global average
value, as the scale of homogeneity is approached at roughly the baryon
acoustic oscillation (BAO) scale of
$110h^{-1}$Mpc. This predicted effect could account for the Hubble bubble
\cite{JRK} and more detailed studies of the scale
dependence of the local Hubble flow \cite{LS}.

In fact, the variance of the local Hubble flow below the scale of homogeneity
should correlate strongly to observed structures in a manner which has no
equivalent prediction in FLRW models.

There is already evidence from the study of large--scale bulk flows that
apparent peculiar velocities determined in the FLRW framework have a
magnitude in excess of the expectations of the standard \LCDM\ model
\cite{WFH,Kash}. In the present framework, rather than having
a uniform expansion (with respect to one set of clocks), with respect to
which peculiar velocities are defined, we have variations in the expansion
rate in regions of different density which are expanding but decelerating
at different rates. Nonetheless, given that our location is on the edge
of a dominant void and a wall \cite{Tully} the equivalent maximum peculiar
velocity can be estimated as
\beq
v\ns{pec}=(\frn32\Hb-H\Z0){30\over h}\w{Mpc}=510^{+210}_{-260}\w{km/s}
\eeq
assuming a diameter of $30\,h^{-1}\,$Mpc for the local dominant void.
This rough estimate is of a magnitude consistent with observation.

\section{Future observational tests}

There are two types of potential cosmological tests that can be developed;
those relating to scales below that of statistical homogeneity as discussed
above, and those
that relate to averages on our past light cone on scales much greater than
the scale of statistical homogeneity. The second class of tests includes
equivalents to all the standard cosmological tests of the standard FLRW model
with Newtonian perturbations. This second class of tests can be
further divided into tests which just deal with the bulk cosmological
averages (luminosity and angular diameter distances etc), and those that
deal with the variance from the growth of structures (late epoch integrated
Sachs--Wolfe effect, cosmic shear, redshift space distortions etc). Here
I will concentrate solely on the simplest tests which are directly related
to luminosity and angular diameter distance measures.

In the \name\ cosmology we have an effective dressed luminosity distance
\beq\dL=a\Z0(1+z)\rw,\eeq where $a\Z0=\gc^{-1}\ab\Z0$, and
\beq\rw=\gb\fvf^{1/3}
\int_\ts^{\ts\X0}{\dd\tb\over\gb(\tb)(1-\fv(\tb))^{1/3}\ab(\tb)}\,.
\label{eq:dL}\eeq
We can also define an {\em effective angular diameter distance}, $\dA$, and an
{\em effective comoving distance}, $D$, to a redshift $z$ in the
standard fashion
\beq\dA={D\over1+z}={\dL\over(1+z)^2}\,.\label{dist}\eeq

\begin{figure}
\centerline{{\sbf(a)}\hskip-15pt
\includegraphics[width=2.25in]{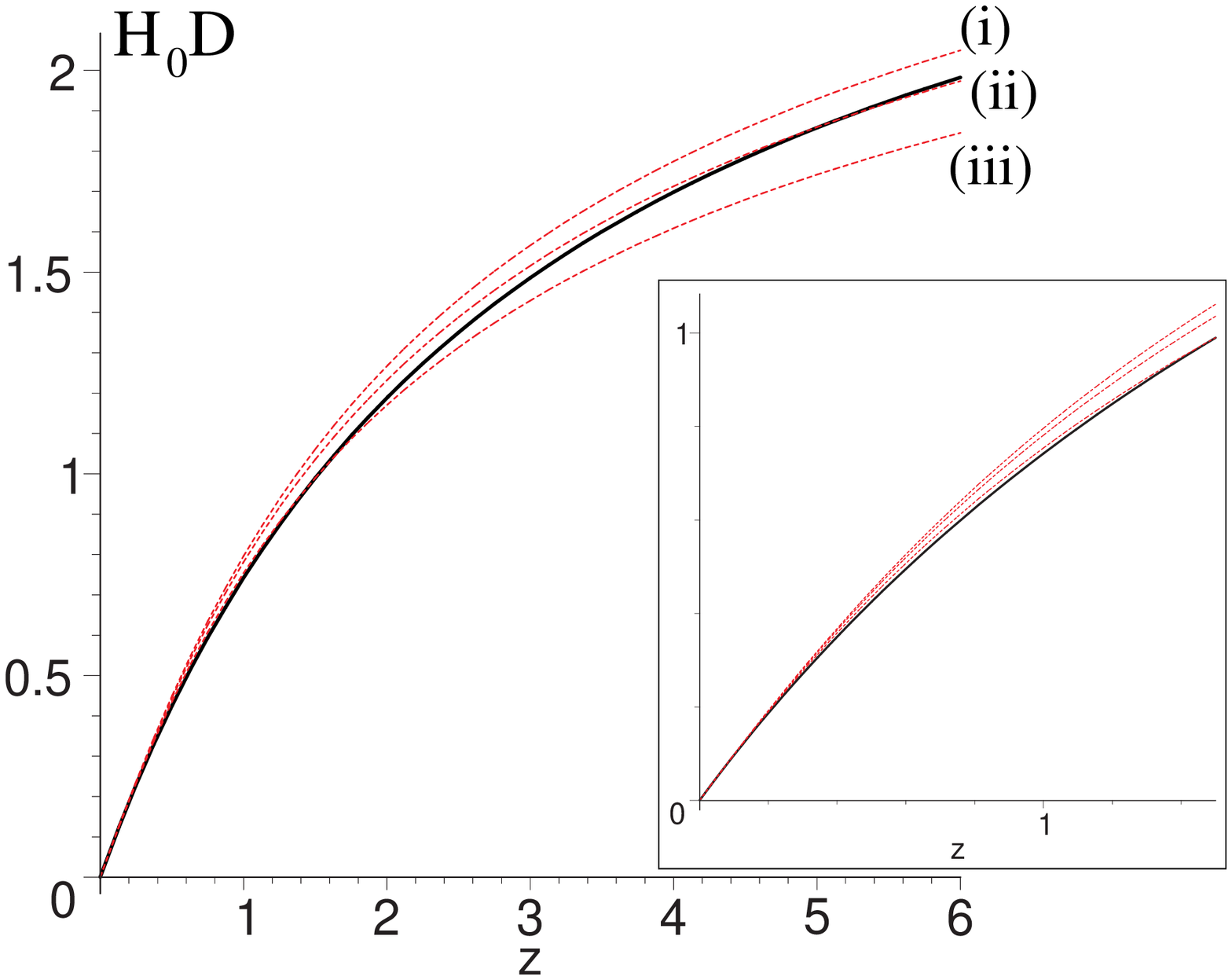}
\qquad\qquad{\sbf(b)}\hskip-15pt
\includegraphics[width=2.in]{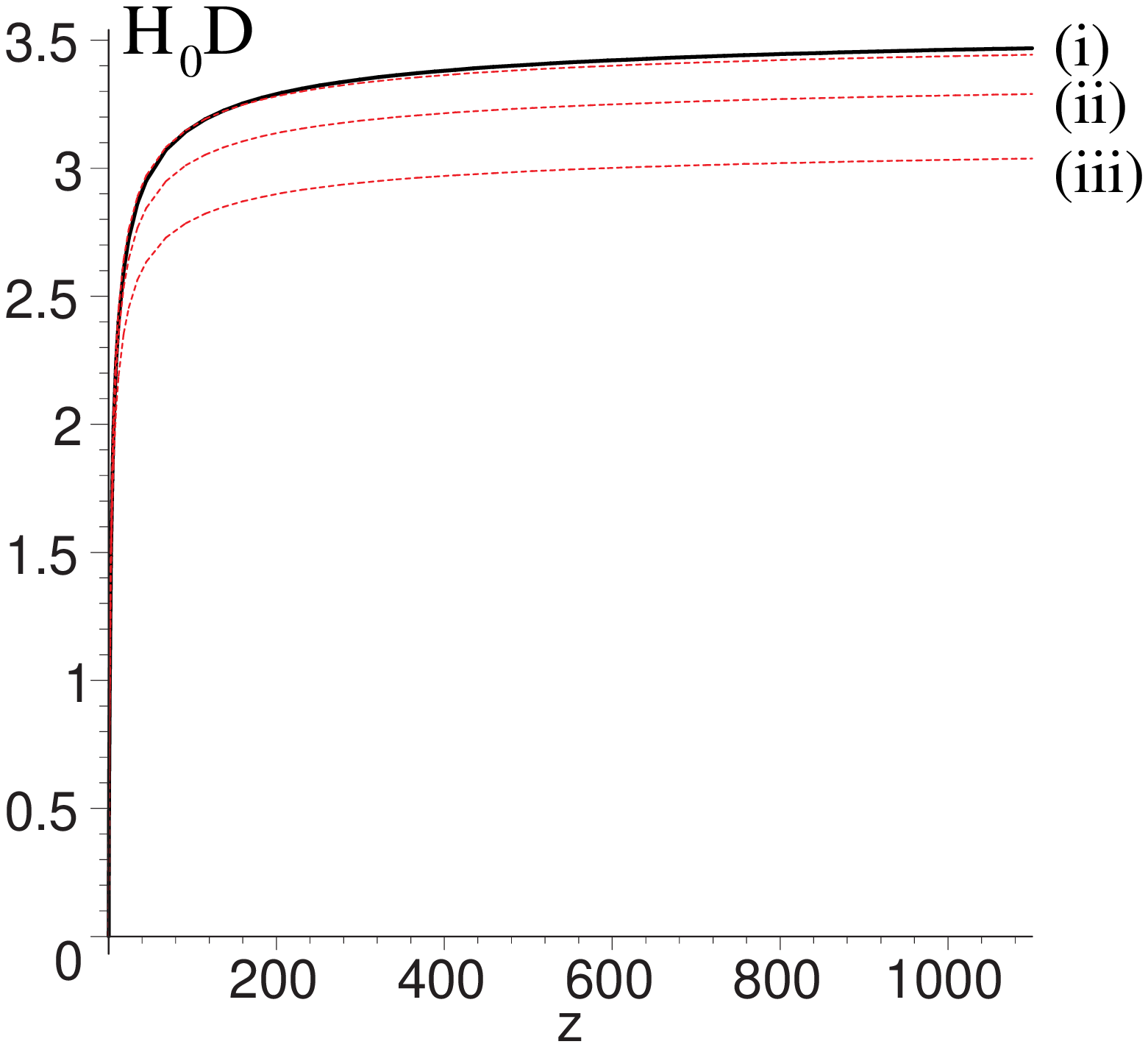}
}
\caption{The effective comoving distance $\Hx D(z)$ is plotted for the
best-fit \name\ (TS) model, with $\fvn=0.762$, (solid line); and for various
spatially flat \LCDM\ models (dashed lines). The parameters for the dashed
lines are (i) $\OmMn=0.249$ (best-fit to WMAP5 only \cite{wmap5}); (ii)
$\OmMn=0.279$ (joint best-fit to SneIa, BAO and WMAP5); (iii) $\OmMn=0.34$
(best-fit to Riess07 SneIa only \cite{Riess07}). Panel {\bf(a)} shows the
redshift range $z<6$, with an inset for $z<1.5$, which is the range tested by
current SneIa data. Panel {\bf(b)} shows the range $z<1100$ up to the
surface of last scattering, tested by WMAP.}
\label{fig_coD}
\end{figure}

A direct method of comparing the distance measures with those of homogeneous
models with dark energy, is to observe that for a standard spatially
flat cosmology with dark energy obeying an equation of state $P\Z D=w(z)
\rh\Z D$, the quantity
\beq
\Hm D=\int_0^z{\dd z'\over \sqrt{\OmMn(1+z')^3+\Omega\Z{D0}
\exp\left[3\int_0^{z'}{(1+w(z''))\dd z''\over 1+z''}\right]}}\,,
\label{rFLRW}\eeq
does not depend on the value of the Hubble constant, $\Hm$, but only
directly on $\OmMn=1-\Omega\Z{D0}$. Since the best-fit values of $\Hm$
are potentially different for the different scenarios, a comparison of
$\Hm D$ curves as a function of redshift for the \name\ model versus the
\LCDM\ model gives a good indication of where the largest differences can be
expected, independently of the value of $\Hm$. Such a
comparison is made in Fig.~\ref{fig_coD}.

We see that as redshift increases the \name\ model interpolates between
\LCDM\ models with different values of $\OmMn$. For redshifts $z\lsim1.5$
$\Dfb$ is very close to $\Dlcdm$ for the parameter values $(\OmMn,\OmLn)
=(0.34,0.66)$ (model (iii)) which best-fit the Riess07 supernovae (SneIa) data
\cite{Riess07} only, by our own analysis. For very large
redshifts that approach the surface of last scattering, $z\lsim1100$, on the
other hand, $\Dfb$ very closely matches $\Dlcdm$ for the parameter values
$(\OmMn,\OmLn) =(0.249,0.751)$ (model (i)) which best-fit WMAP5 only
\cite{wmap5}. Over redshifts $2\lsim z\lsim10$, at which scales independent
tests are conceivable, $\Dfb$ makes a transition over corresponding
curves of $\Dlcdm$ with intermediate values of $(\OmMn,\OmLn)$. The
$\Dlcdm$ curve for joint best-fit parameters to SneIa, BAO measurements and
WMAP5 \cite{wmap5}, $(\OmMn,\OmLn) =(0.279,0.721)$ is best--matched over
the range $5\lsim z\lsim 6$, for example.

The difference of $\Dfb$ from any single $\Dlcdm$ curve is perhaps most
pronounced in the range $2\lsim z\lsim 6$, which may be an optimal regime to
probe in future experiments. Gamma--ray bursters (GRBs) now probe distances to
redshifts $z\lsim8.3$, and could be very useful if their properties could be
understood to the extent that they might be reliably used as standard candles.
A considerable amount work of work has already been done on Hubble diagrams for
GRBs. (See, e.g., \cite{GRB}.) Much more work is needed to nail down systematic
uncertainties, but GRBs may eventually provide a definitive test in future. An
analysis of the \name\ model Hubble diagram using 69 GRBs has just been
performed by Schaefer \cite{Schaefer}, who finds that the \name\ model fits
the data better than the concordance \LCDM\ model, but not yet by a huge
margin\footnote{By contrast the conformal gravity model of Mannheim \cite{mann}
produced a worse fit, while the Chaplyagin gas fit best only in the limit that
its parameters reduce to those of the \LCDM\ model \cite{Schaefer}.}. As more
data is accumulated, it should become possible to distinguish the models if
the issues with the standardization of GRBs can be ironed out.

\subsection{The effective ``equation of state''}

The shape of the $\Hm D$ curves depicted in
Fig.~\ref{fig_coD} represents the observable quantity one is actually
measuring in tests some researchers loosely refer to as ``measuring the
equation of state''. For spatially flat dark energy models, with $\Hm D$
given by (\ref{rFLRW}), one finds that the function $w(z)$ appearing in the
fluid equation of state $P\Z D=w(z)\rh\Z D$ is related to the first
and second derivatives of (\ref{rFLRW}) by
\beq
w(z)={\frn23(1+z)D'^{-1}D''+1\over\OmMn(1+z)^3\Hm^2 D'^2-1}
\label{eos}\eeq
where prime denotes a derivative with respect to $z$. Such a relation
can be applied to observed distance measurements, regardless of whether
the underlying cosmology has dark energy or not. Since it involves
first and second derivatives of the observed quantities, it is actually
much more difficult to determine observationally than directly fitting
$\Hm D(z)$.

\begin{figure}
\centerline{{\sbf(a)}\hskip-5pt
\includegraphics[width=2.in]{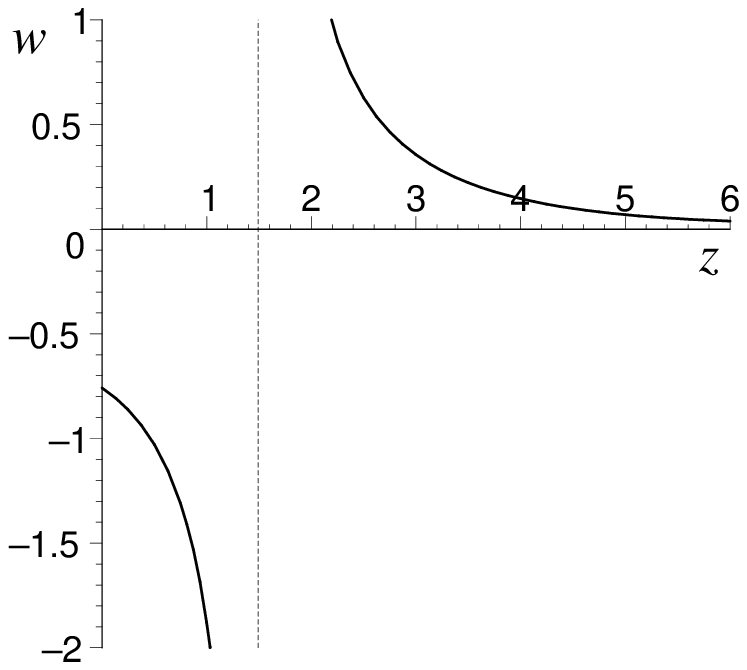}\qquad\qquad
{\sbf(b)}\hskip-5pt
\includegraphics[width=2.in]{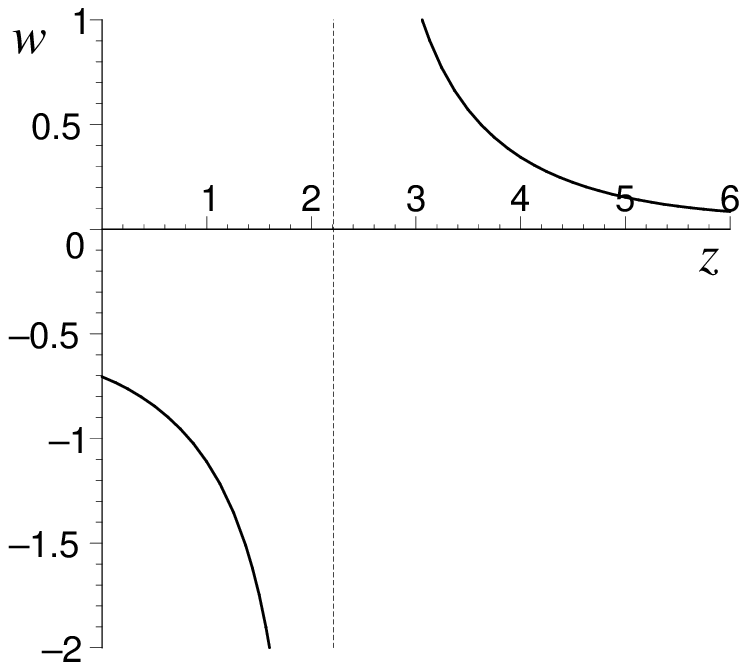}
}
\caption{The artificial equivalent of an equation of state
constructed using the effective comoving distance
(\ref{eos}), plotted for the \name\ tracker solution with best-fit value
$\fvn=0.762$, and two different values of $\OmMn$: {\bf(a)} the canonical
dressed value $\OmMn=\frn12(1-\fvn)(2+\fvn)=0.33$; {\bf(b)} $\OmMn=0.279$.}
\label{fig_wz}
\end{figure}

The equivalent of the ``equation of state'', $w(z)$, for the \name\ model
is plotted in Fig.~\ref{fig_wz}. The fact that $w(z)$ is undefined at
a particular redshift and changes sign through $\pm\infty$ simply reflects
the fact that in (\ref{eos}) we are dividing by a quantity which goes
to zero for the \name\ model, even though the underlying curve of
Fig.~\ref{fig_coD} is smooth. Since one is not dealing with a dark energy
fluid in the present case, $w(z)$ simply has no physical meaning.
Nonetheless, phenomenologically the results
do agree with the usual inferences about $w(z)$ for fits of standard dark
energy cosmologies to SneIa data. For the canonical model of
Fig.~\ref{fig_wz}(a) one finds that the average value of $w(z)\simeq-1$
on the range $z\lsim0.7$, while the average value of $w(z)<-1$ if the
range of redshifts is extended to higher values. The $w=-1$ ``phantom divide''
is crossed at $z\simeq0.46$ for $\fvn\simeq0.76$. One recent study \cite{ZZ}
finds mild 95\% evidence for an equation of state that crosses the phantom
divide from $w>-1$ to $w<-1$ in the range $0.25<z<0.75$ in accord with
the \name\ expectation. By contrast,
another study \cite{SCHMPS} at redshifts $z<1$ draws different conclusions
about dynamical dark energy, but for the given uncertainties in $w(z)$ the
data is consistent with Fig.~\ref{fig_coD}(a) as well as with a cosmological
constant \cite{obs}.

The fact that $w(z)$
is a different sign to the dark energy case for $z>2$ is another
way of viewing our statement above that the redshift range $2\lsim z\lsim6$
may be optimal for discriminating model differences.
\subsection{The $H(z)$ measure}

Further observational diagnostics can be devised if the expansion rate
$H(z)$ can be observationally determined as a function of redshift. Recently
such a determination of $H(z)$ at $z=0.24$ and $z=0.43$ has been made using
redshift space distortions of the BAO scale
in the \LCDM\ model \cite{GCH}. This technique is
of course model dependent, and the Kaiser effect would have to be re-examined
in the \name\ model before a direct comparison of observational results could
be made. A model--independent measure of $H(z)$, the redshift time drift
test, is discussed below.

\begin{figure}
\centerline{\includegraphics[width=2.in]{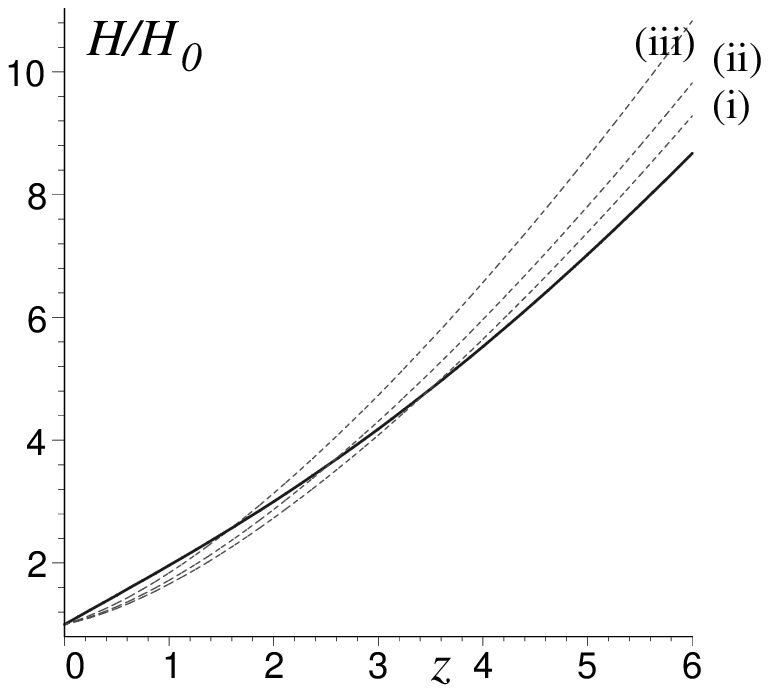}}
\caption{The function $\Hm^{-1} H(z)$ for the \name\ model with
$\fvn=0.762$ (solid line) is compared to $\Hm^{-1} H(z)$ for three
spatially flat \LCDM\ models with the same values of $(\OmMn,\OmLn)$ as in
Fig.~\ref{fig_coD} (dashed lines).}
\label{fig_HH0}
\end{figure}

In Fig.~\ref{fig_HH0} we compare $H(z)/\Hm$ for the \name\ model to
spatially flat \LCDM\ models with the same parameters chosen in
Fig.~\ref{fig_coD}. The most notable feature is that the slope of $H(z)/\Hm$
is less than in the \LCDM\ case, as is to be expected for a model whose
(dressed) deceleration parameter varies more slowly than for \LCDM.
\subsection{The $Om(z)$ measure}

Recently a number of authors \cite{GCC,SSS,ZC} have discussed various
roughly equivalent diagnostics of dark energy. For example, Sahni, Shafieloo
and Starobinsky \cite{SSS}, have proposed a diagnostic function
\beq
Om(z)=\Bigl[{H^2(z)\over\Hm^2}-1\Bigr]\left[(1+z)^3-1\right]^{-1}\,,
\label{dSSS}\eeq
on account of the fact that it is equal to the constant present epoch
matter density parameter, $\OmMn$, at all redshifts for a spatially flat
FLRW model with pressureless dust and a cosmological constant. However, it is
not constant if the cosmological constant is replaced by other forms of dark
energy. For general FLRW models, $H(z)=[D'(z)]^{-1}\sqrt{1+\Omkn\Hm^2 D^2(z)}$,
which only involves a single derivatives of $D(z)$. Thus the diagnostic
(\ref{dSSS}) is easier to reconstruct observationally than the equation
of state parameter, $w(z)$.
\begin{figure}[htb]
\centerline{{\sbf(a)}\hskip-5pt
\includegraphics[width=2.2in]{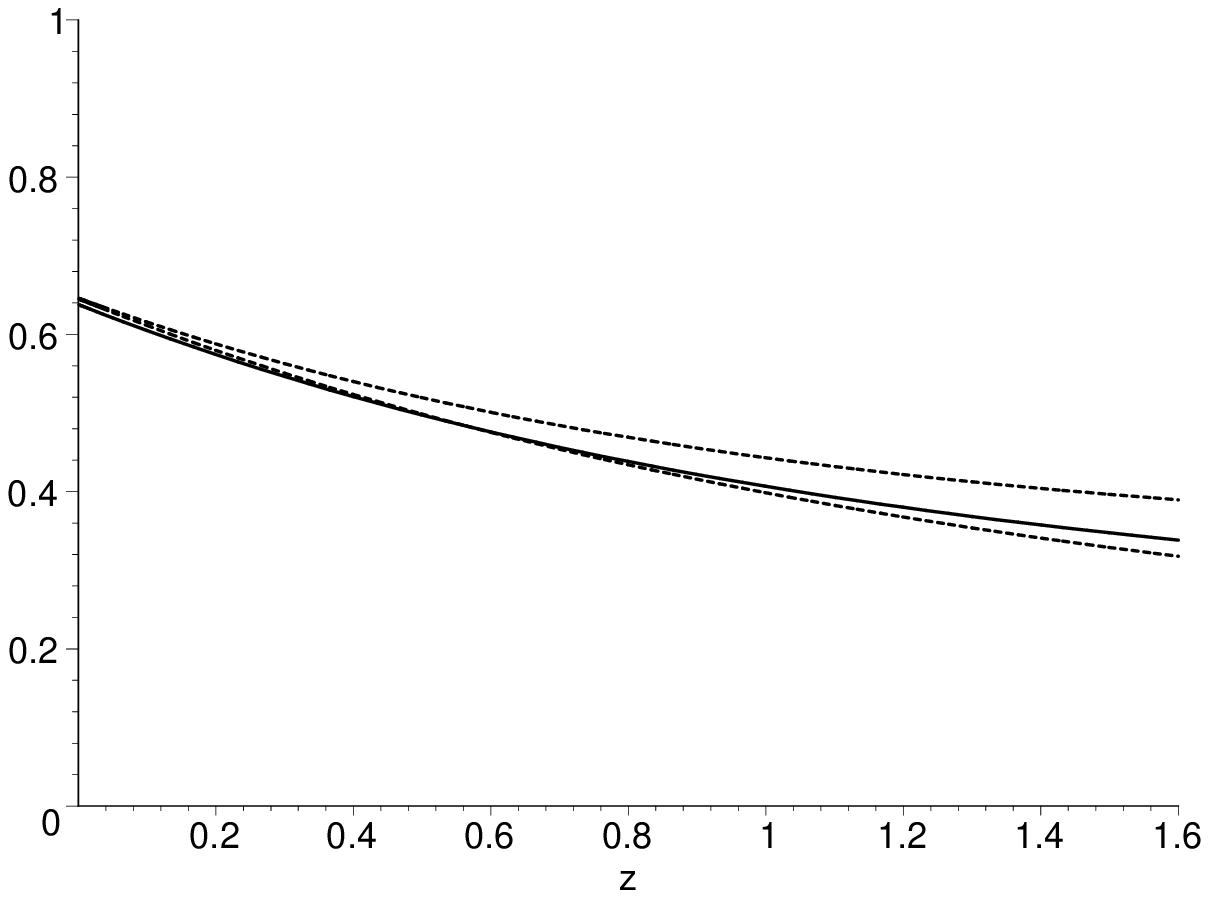}
\qquad
{\sbf(b)}\hskip-5pt
\includegraphics[width=2.2in]{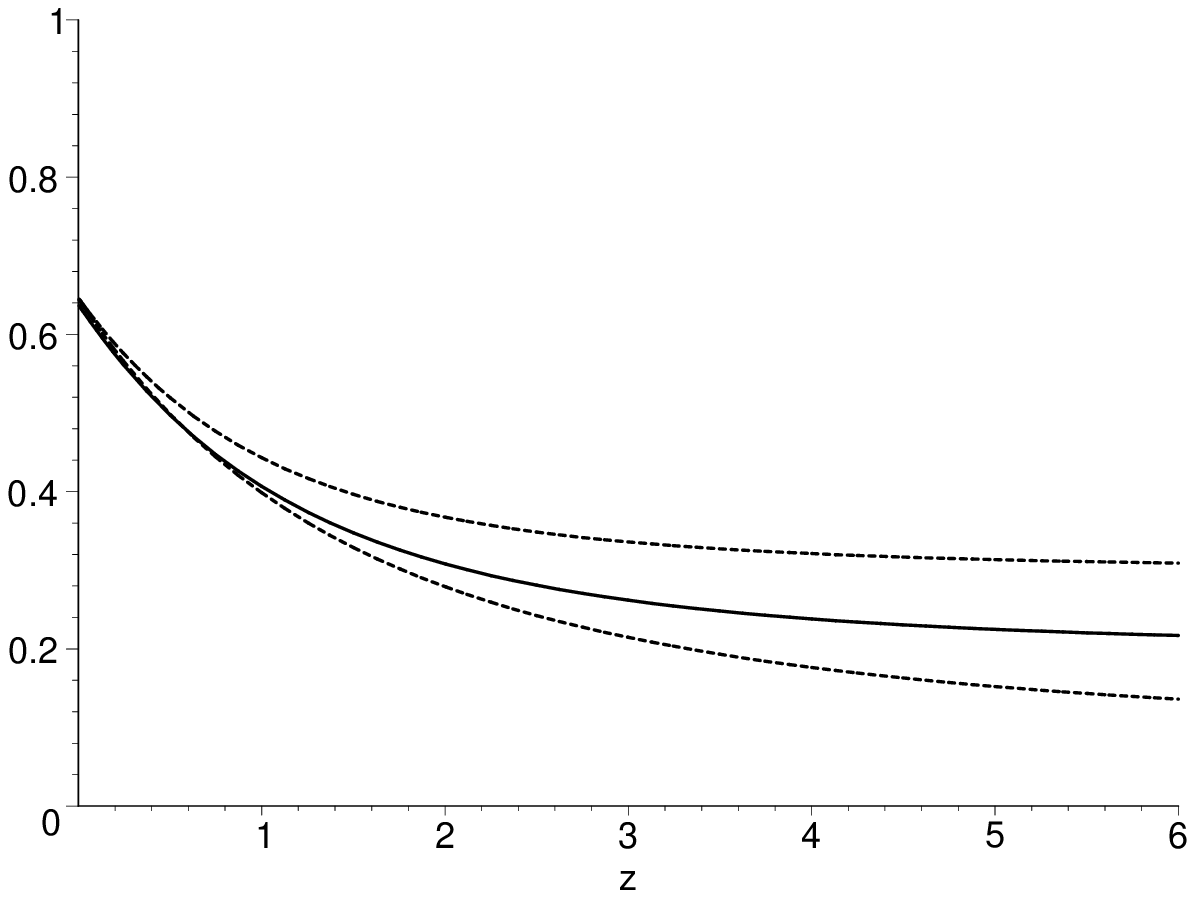}
}
\caption{The dark energy diagnostic $Om(z)$ of Sahni, Shafieloo and Starobinsky
\cite{SSS} plotted for the \name\ tracker solution with best-fit value
$\fvn=0.762$ (solid line), and $1\si$ limits (dashed lines) from ref.\
\cite{LNW}: {\bf(a)} for the redshift range $0<z<1.6$ as shown in
ref.\ \cite{SSS2}; {\bf(b)} for the redshift range $0<z<6$.}
\label{fig_Om}
\end{figure}

The quantity $Om(z)$ is readily calculated for the \name\ model, and the
result is displayed in Fig.~\ref{fig_Om}. What is
striking about Fig.~\ref{fig_Om}, as compared to the curves for quintessence
and phantom dark energy models as plotted in ref.\ \cite{SSS}, is that the
initial value
\beq
Om(0)=\frn23\left.H'\right|_0={2(8\fvn^3-3\fvn^2+4)(2+\fvn)\over(4\fvn^2+\fvn
+4)^2}
\label{intc}\eeq
is substantially larger than in the spatially flat dark energy models.
Furthermore, for the \name\ model $Om(z)$ does not asymptote to the dressed
density parameter $\OmMn$ in any redshift range. For quintessence models
$Om(z)>\OmMn$,
while for phantom models $Om(z)<\OmMn$, and in both cases $Om(z)\to\OmMn$
as $z\to\infty$. In the \name\ model, $Om(z)>\OmMn\simeq0.33$ for $z\lsim1.7$,
while $Om(z)<\OmMn$ for $z\gsim1.7$. It thus behaves more like a quintessence
model for low $z$, in accordance with Fig.~\ref{fig_wz}. However, the
steeper slope and the different large $z$ behaviour mean the
diagnostic is generally very different to that of typical dark energy models.
For large $z$, $\OMMn<Om(\infty)<\OmMn$, if $\fvn>0.25$.

Interestingly enough, a recent analysis of SneIa, BAO and CMB data
\cite{SSS2} for dark energy models
with two different empirical fitting functions for $w(z)$ gives an intercept
$Om(0)$ which is larger than expected for typical quintessence or phantom
energy models, and in the better fit of the two models the intercept (see
Fig.~3 of ref.\ \cite{SSS2})] is close to the value expected for the \name\
model, which is tightly constrained to the range $0.638<Om(0)<0.646$ if
$\fvn=0.76^{+0.12}_{-0.09}$.

\subsection{The Alcock--Paczy\'nski test and baryon acoustic oscillations}

Some time ago Alcock and Paczy\'nski devised a test \cite{AP} which relies on
comparing the radial and transverse proper length scales of spherical standard
volumes comoving with the Hubble flow. This test, which determines the function
\beq
f\Ns{AP}={1\over z}\left|\Deriv\de z\th\right|={HD\over z},
\label{fAP}
\eeq
was originally conceived to distinguish FLRW models with a cosmological
constant from those without a $\Lambda$ term. The test is free from many
evolutionary effects, but relies on one being able to remove systematic
distortions due to peculiar velocities.

\begin{figure}
\centerline{{\sbf(a)}\hskip-5pt
\includegraphics[width=1.8in]{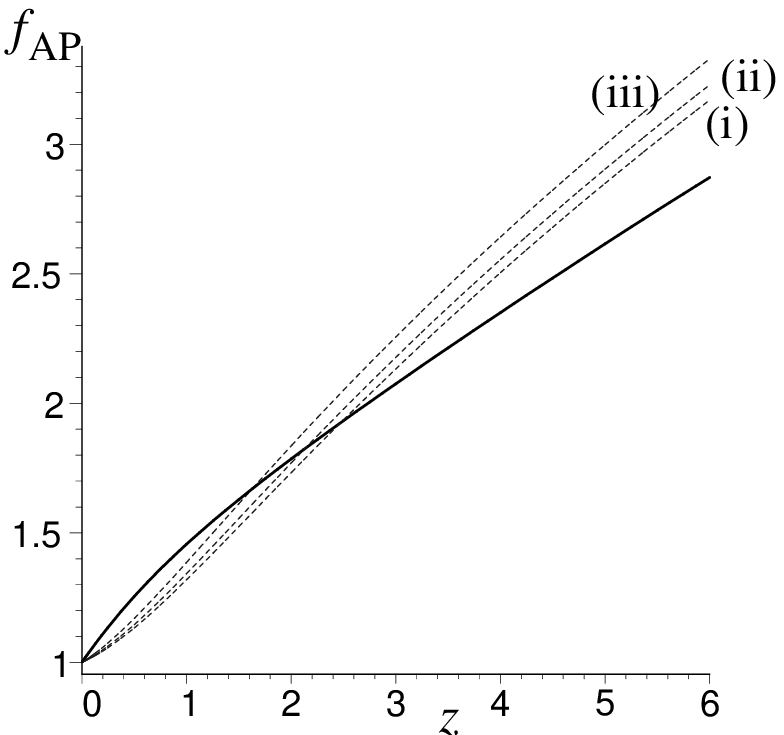}
\qquad\qquad
{\sbf(b)}\hskip-5pt
\includegraphics[width=2.in]{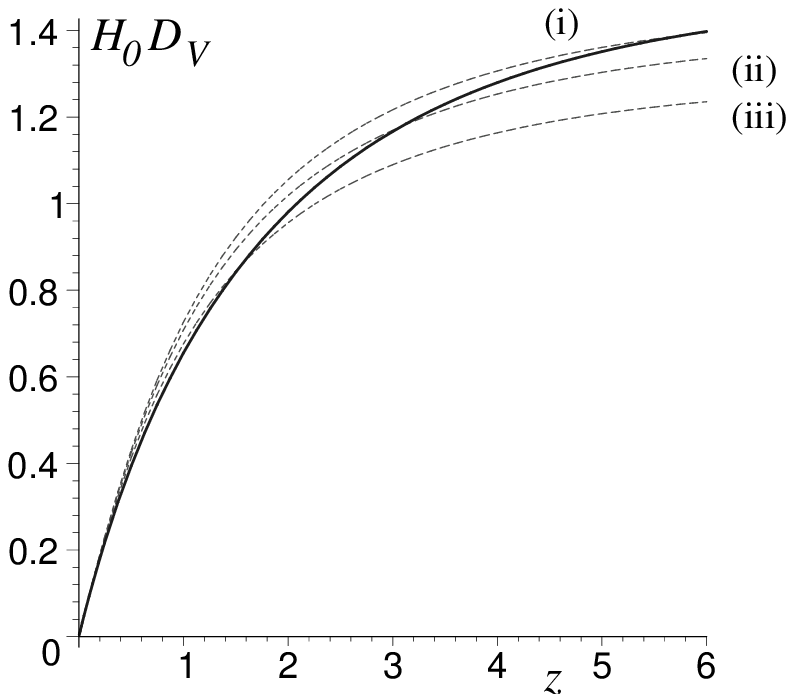}
}
\caption{{\bf(a)} The Alcock--Paczy\'nski test function $f\Ns{AP}=HD/z$;
and {\bf(b)} the BAO radial test function $\Hm D\Z V=\Hm Df\Ns{AP}^{-1/3}$.
In each case the \name\ model with
$\fvn=0.762$ (solid line) is compared to three
spatially flat \LCDM\ models with the same values of $(\OmMn,\OmLn)$ as in
Fig.~\ref{fig_coD} (dashed lines).}
\label{fig_AP}
\end{figure}
Current detections of the BAO scale in galaxy clustering statistics
\cite{bao,Percival} can in fact be viewed as a variant of the
Alcock--Paczy\'nski test, as they make use of both the transverse and
radial dilations of the fiducial comoving BAO scale to present a measure
\beq
D\Z V=\left[zD^2\over H(z)\right]^{1/3}=Df\Ns{AP}^{-1/3}.
\label{BAOr}\eeq

In Fig.~\ref{fig_AP} the Alcock--Paczy\'nski test function (\ref{fAP})
and BAO scale measure (\ref{BAOr}) of the \name\ model are
compared to those of the spatially flat \LCDM\ model with different values of
($\OmLn$,$\OmLn$). Over the range of redshifts $z<1$ studied currently
with galaxy clustering statistics, the $f\Ns{AP}$ curve distinguishes
the \name\ model from the \LCDM\ models much more strongly than the
$D\Z V$ test function. In particular, the \name\ $f\Ns{AP}$ has a distinctly
different shape to that of the \LCDM\ model, being convex. The primary
reason for use of the integral measure (\ref{BAOr}) has been a lack of data.
Future measurements with enough data to separate the radial and angular BAO
scales are a potentially powerful way of distinguishing the \name\ model from
\LCDM.

Recently Gazta\~naga, Cabr\'e and Hui \cite{GCH} have made the first
efforts to separate the radial and angular BAO scales in different redshift
slices. Although they have not yet published separate values for the radial
and angular scales, their results are interesting when compared to the
expectations of the \name\ model. Their study yields best-fit values
of the present total matter and baryonic matter density parameters,
$\OmMn$ and $\OmBn$, which are in tension with WMAP5 parameters fit to
the \LCDM\ model. In particular, the ratio of nonbaryonic cold dark
matter to baryonic matter has a best-fit value $\OmCn/\OmBn=(\OmMn-\OmBn)/
\OmBn$ of 3.7 in the $0.15<z<0.3$ sample, 2.6 in the $0.4<z<0.47$ sample,
and 3.6 in the whole sample, as compared to the expected value of 6.1 from
WMAP5. The analysis of the 3--point correlation function yields similar
conclusions, with a best fit \cite{GCCCF} $\OmMn=0.28\pm0.05$,
$\OmBn=0.079\pm0.025$. By comparison, the parameter fit to the \name\ model of
ref.\ \cite{LNW} yields dressed parameters $\OmMn=0.33^{+0.11}_{-0.16}$,
$\OmBn=0.080^{+0.021}_{-0.013}$, and a ratio $\OmCn/\OmBn=3.1^{+2.5}_{-2.4}$.
Since homogeneous dark energy models are not generally expected to give rise
to a renormalization of the ratio of nonbaryonic to baryonic matter, this
is encouraging for the \name\ model.

\subsection{Test of (in)homogeneity}

Recently Clarkson, Bassett and Lu \cite{CBL} have constructed what they call
a ``test of the Copernican principle'' based on the observation that
for homogeneous, isotropic models which obey the Friedmann equation,
the present epoch curvature parameter, a constant, may be written as
\beq
\Omkn={[H(z)D'(z)]^2-1\over[\Hm D(z)]^2}\label{ctest1}
\eeq
for all $z$, irrespective of the dark energy model or any other model
parameters. Consequently, taking a further derivative, the quantity
\beq
\CC(z)\equiv1+H^2(DD''-D'^2)+HH'DD'\label{ctest2}
\eeq
must be zero for all redshifts for any FLRW geometry.

A deviation of $\CC(z)$ from zero, or of (\ref{ctest1}) from a constant
value, would therefore mean that the assumption of homogeneity is violated.
Although this only constitutes a test of the assumption of the Friedmann
equation, i.e., of the Cosmological Principle rather than the broader
Copernican Principle adopted in ref.\ \cite{clocks}, the average
inhomogeneity will give a clear and distinct prediction of a nonzero
$\CC(z)$ for the \name\ model.
\begin{figure}
\centerline{{\sbf(a)}
\includegraphics[width=2.2in]{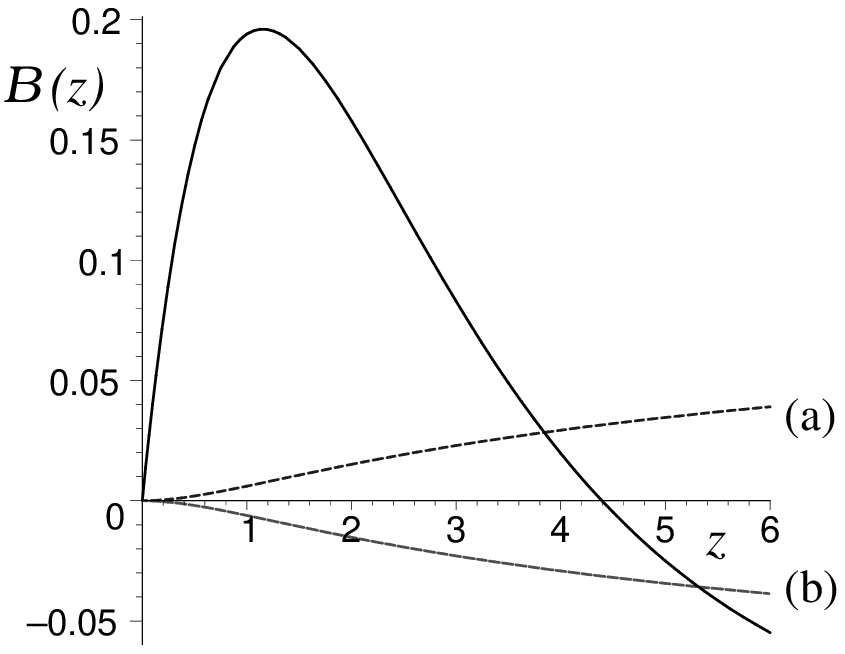}
\qquad
{\sbf(b)}
\includegraphics[width=2.in]{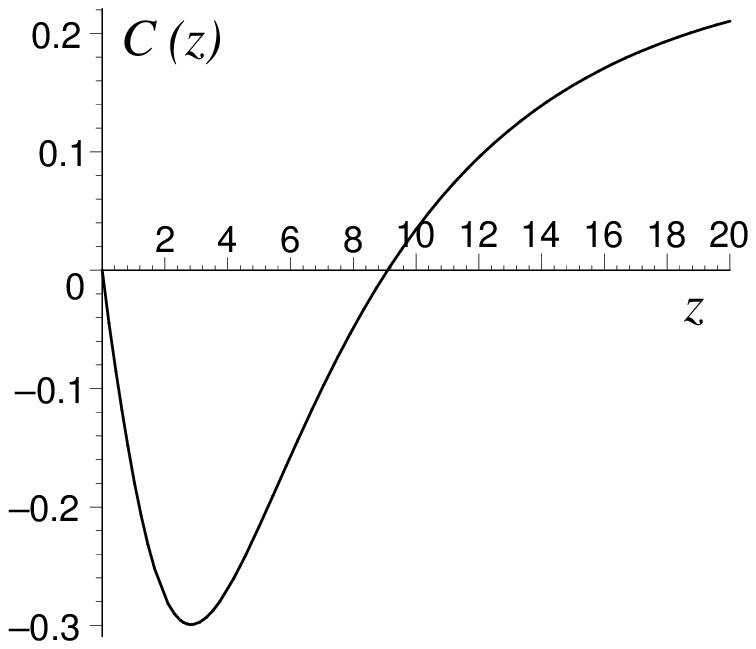}}
\caption{{\bf Left panel:}
The (in)homogeneity test function $\BB(z)=[HD']^2-1$ is plotted for
the \name\ tracker solution with best-fit value $\fvn=0.762$ (solid line), and
compared to the equivalent curves $\BB=\Omkn(\Hm D)^2$ for two different
\LCDM\ models with small curvature:\break {\bf(a)} $\OmMn=0.28$, $\OmLn=0.71$,
$\Omkn=0.01$; {\bf(b)} $\OmMn=0.28$, $\OmLn=0.73$, $\Omkn=-0.01$.\hfil\break
{\bf Right panel:} The (in)homogeneity test function $\CC(z)$ is plotted for
the $\fvn=0.762$ tracker solution.}
\label{fig_Bex}
\end{figure}

The functions (\ref{ctest1}) and (\ref{ctest2}) are computed in ref.\
\cite{obs}.
Observationally it is more feasible to fit (\ref{ctest1}) which involves one
derivative less of redshift. In Fig.\ \ref{fig_Bex} we exhibit both
$\CC(z)$, and also the function
$\BB(z)=[HD']^2-1$ from the numerator of (\ref{ctest1}) for the \name\
model, as compared to two \LCDM\ models with a small amount of spatial
curvature. A spatially flat FLRW model would have $\BB(z)\equiv0$. In other
FLRW cases $\BB(z)$ is always a monotonic
function whose sign is determined by that of $\Omkn$. An open $\Lambda=0$
universe with the same $\OmMn$ would have a monotonic function $\BB(z)$
very much greater than that of the \name\ model.

\subsection{Time drift of cosmological redshifts}

For the purpose of the $Om(z)$ and (in)homogeneity tests considered in the
last section, $H(z)$ must be observationally determined, and this is
difficult to achieve in a model-independent way. There is one way of
achieving this, however, namely by measuring the time variation of the
redshifts of different sources over a sufficiently long time interval
\cite{SML}, as has been discussed recently by Uzan, Clarkson and Ellis
\cite{UCE}. Although the measurement is extremely challenging, it may be
feasible over a 20 year period by precision measurements of the Lyman-$\al$
forest in the redshift range $2<z<5$ with the next generation of
Extremely Large Telescopes \cite{ELT}.

\begin{figure}[htb]
\centerline{\includegraphics[width=2.2in]{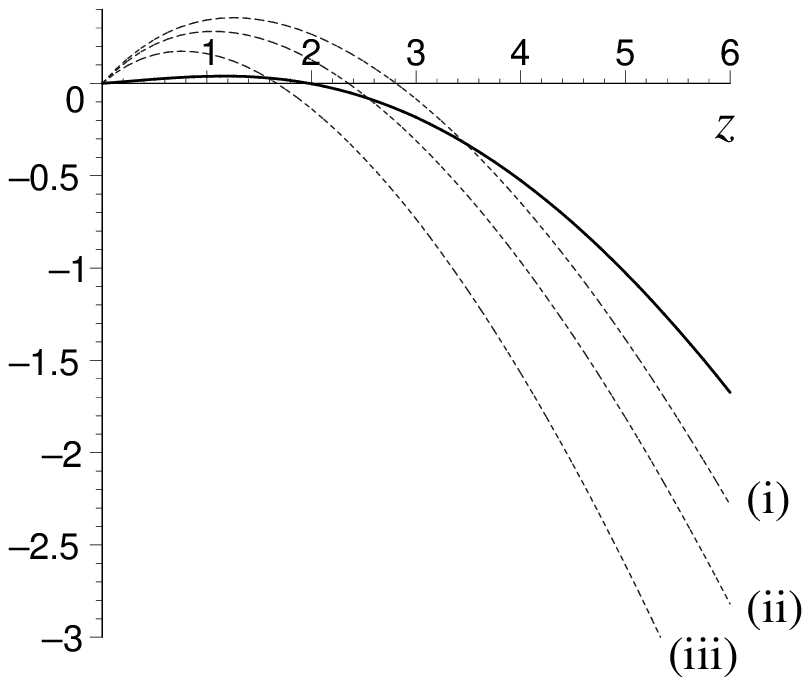}}
\caption{The function $\Hm^{-1}\Deriv\dd\ta z$ for the \name\ model with
$\fvn=0.762$ (solid line) is compared to $\Hm^{-1}\Deriv\dd\ta z$ for three
spatially flat \LCDM\ models with the same values of $(\OmMn,\OmLn)$ as in
Fig.~\ref{fig_coD} (dashed lines).}
\label{fig_zdot}
\end{figure}

In ref.\ \cite{obs} an analytic expression for $\Hm^{-1}\Deriv\dd\ta z$
is determined, the derivative being with respect to wall time for observers
in galaxies. The resulting function is displayed in Fig.~\ref{fig_zdot} for
the best-fit \name\ model with $\fvn=0.762$, where it is compared to the
equivalent function for three different spatially flat \LCDM\ models. What
is notable is that the curve for the \name\ model is considerably flatter
than those of the \LCDM\ models. This may be understood to arise from
the fact that the magnitude of the apparent acceleration is considerably
smaller in the \name\ model, as compared to the magnitude of the acceleration
in \LCDM\ models. For models in which there is no apparent acceleration
whatsoever, one finds that $\Hm^{-1}\Deriv\dd\ta z$ is always negative.
If there is cosmic acceleration at late epochs, real or apparent, then
$\Hm^{-1}\Deriv\dd\ta z$ will become positive at low redshifts, though
at a somewhat larger redshift than that at which acceleration is deemed
to have begun.

Fig.~\ref{fig_zdot} demonstrates that a very clear signal of differences in
the redshift time drift between the \name\ model and \LCDM\ models might
be determined at low redshifts when $\Hm^{-1}\Deriv\dd\ta z$ should be
positive. In particular, the magnitude of $\Hm^{-1}\Deriv\dd\ta z$ is
considerably smaller for the \name\ model as compared to \LCDM\ models.
Observationally, however, it is expected that measurements
will be best determined for sources in the Lyman $\al$ forest in the range,
$2<z<5$. At such redshifts the magnitude of the drift is somewhat
more pronounced in the case of the \LCDM\ models. For a source at $z=4$,
over a period of $\de\ta=10$ years we would have $\de z=-3.3\times10^{-10}$
for the \name\ model with $\fvn=0.762$ and $\Hm=61.7\kmsMpc$. By comparison,
for a spatially flat \LCDM\ model with $\Hm=70.5\kmsMpc$ a source
at $z=4$ would over ten years give $\de z=-4.7\times10^{-10}$ for
$(\OmMn,\OmLn)=(0.249,0.751)$, and $\de z=-7.0\times10^{-10}$ for
$(\OmMn,\OmLn)=(0.279,0.721)$.
\section{Dark matter and the \name\ scenario}

Since much of this conference has been about alternatives to standard
nonbaryonic dark matter, I will briefly comment on the issue of dark matter
vis--\`a--vis the \name\ scenario. The \name\ model only addresses
large scale cosmological averages, and does not make specific
predictions about dark matter typically inferred from direct observations
of bound systems, such as rotation curves of galaxies, gravitational lensing
or motions of galaxies within clusters. However, as has been discussed
in ref.\ \cite{clocks}, in a re-examination of the post-Newtonian
approximation within the averaging problem in cosmology,
departures from the na\"{\i}ve Newtonian limit in an asymptotically flat
space are to be expected. Consequently, any approach to treat galactic
dynamics as a fully nonlinear problem within general relativity, such as the
work discussed by Cooperstock at this conference \cite{CT1,CT2}, is to be
expected as potentially viable and compatible with the \name\ scenario.

As discussed above, in the \name\ scenario fits to supernova data,
the BAO scale and the angular diameter distance of the sound horizon
in CMB anisotropy data allow one to estimate the dressed matter density
parameter, $\OmMn$, while primordial nucleosynthesis bounds allow us to
independently estimate the corresponding baryonic matter density
parameter, $\OmBn$, and consequently of the nonbaryonic matter density
parameter $\OmCn\equiv\OmMn-\OmBn$. As a result we find a mass ratio of
nonbaryonic dark matter to baryonic matter of $\OmCn/\OmBn=3.1^{+2.5}
_{-2.4}$, with uncertainties from supernova data alone, or tighter bounds if
constraints based on the angular diameter distance to the sound horizon are
imposed. This potentially reduces the relative amount of nonbaryonic matter
by a factor of two or more as compared to the standard concordance cosmology.

The Cooperstock--Tieu model \cite{CT1,CT2} demonstrates that the rotation
curves of spiral galaxies can be reproduced\footnote{See ref.\ \cite{CC} for
recent work which increases the number of galaxies whose rotation curves have
been successfully fit by this model.} by a stationary axisymmetric
rotating dust solution, obviating the need for spherical halos of dark matter
as a required by the na\"{\i}ve use of Newtonian dynamics\footnote{A number of
details of the \CT\ model have been disputed; see, e.g., ref.\
\cite{RS} for a summary. While the details are open to debate \cite{CT2,RS},
any deficiencies of the model might easily be an artefact of the
simplifications of not including gas pressure, or other realistic features such
as differentially rotating spiral arms and central bars. The basic premise of
the \CT\ model, namely that nonlinearities in the Einstein equations can be
important even in the weak field regime, stands as a consequence of general
relativity that must be seriously considered at the galactic level.}. However,
the \CT\ model does not specify the particle content of the dust. For
the Milky Way the \CT\ mass estimate\footnote{A recent Newtonian estimate of
the mass of the Milky Way \cite{GBGK}, including its dark halo, gives a mass
in the range ($5.7$ -- $10)\times10^{11}\Msun$ within a radius $80\,$kpc, or a
total virial mass of $(1.6\pm0.3)\times 10^{12}\Msun$ at the virial radius
$R\ns{vir}=300\,$kpc. A direct numerical comparison with the \CT\ mass estimate
is difficult, as the latter is confined to the mass within $30\,$kpc of the
galaxy centre for which rotation curve data was available, and different
density profiles can be assumed in the dark outer regions. However, by any
measure the \CT\ mass estimate is certainly considerably reduced relative to
the Newtonian estimate based on a dark matter halo.} of $2.1\times10^{11}\Msun$
\cite{CT2} is a factor of 3--5 times {\em larger} than direct estimates of the
combined baryonic mass of the galactic disk, bulge, bar and nucleus which are
in the range\footnote{The central supermassive black hole, of mass $\goesas4
\times10^9\Msun$ is not included in this estimate of the Milky Way baryonic
mass, as it is omitted in the \CT\ model.}\break ($4.2$ -- $7.2)\times10^{10}
\Msun$ \cite{KZS}. The ratio of the \CT\ mass to the observed baryonic mass
of the Milky Way is thus in agreement with the global timescape estimate
$\OmMn/\OmBn=4.1^{+2.5}_{-2.4}$. Consequently, it is certainly possible for
a significant amount of nonbaryonic dark matter to exist within the universe,
reduced relative to the Newtonian dynamics estimate, if the \CT\ model, or
something close to it\footnote{Another variant of the \CT\ model has been
explored by Balasin and Grumiller \cite{BG}.}, operates at the galactic level.

The fact that several modified gravity approaches, including MOND \cite{mond},
MOG \cite{mog}, and conformal gravity \cite{mann}, are able to
phenomenologically reproduce various aspects of galactic and galaxy cluster
dynamics to varying degrees of success, suggests that some simplifying
principle remains to be found despite the amazing variety of structures
described, which are far too complex to be modelled by simple exact dust
solutions of general relativity such as those of refs.\ \cite{CT1,CT2}. Given
a number of models which fit the same data \cite{bro}, what is needed is that
other falsifiable predictions of all the models are developed, which might rule
some out. The standard Newtonian CDM hypothesis is at least so well developed
that it leads to several testable predictions about structure formation; to the
extent that it is arguably {\em already ruled out} by detailed observational
studies of Local Group galaxies \cite{Kroupa}.

Although na\"{\i}vely the \name\ model appears to predict of order
three times as much nonbaryonic matter as baryonic matter by mass, we should
be careful to note that the ratio $\OmMn/\OmBn$ measures the density
of clumped matter at the present epoch relative to that of baryonic matter
inferred from primordial nucleosynthesis, only when using solutions {\em
averaged on cosmological scales}. Given the problems of defining gravitational
energy in the absence of a timelike Killing vector or other exact symmetries,
we should remain open to the possibility that the nature of clumped
gravitational mass in cosmological averages is more than simply the sum of its
particle constituents. Thus the difference, $\OmMn-\OmBn$, might not simply
be nonbaryonic dark matter particles, but could include some component
of gravitational energy that enters on some relevant scale of coarse-graining
of bound systems, such as the transition from individual galaxies to galaxy
clusters. Even for individual galaxies, such the Milky Way, the difference
between the \CT\ mass estimate and the observed baryonic mass estimate could
either simply be unaccounted dark matter (baryonic or nonbaryonic), or else
include at least a partial contribution from gravitational energy that enters
in the coarse--graining of the dust. Thus we should keep an open mind about
the existence or nonexistence of nonbaryonic dark matter
as long as these questions are not understood.

It is worth mentioning, however, that in the standard cosmology nonbaryonic
dark matter is required to start structure formation going. At the surface of
last scattering, dark matter density contrasts $\de\rh\Z C/\rh\Z C$ are
expected to be an order of magnitude stronger than the baryonic density
contrasts $\de\rh\Z B/\rh\Z B\goesas10^{-5}$. The nonbaryonic dark matter
overdensity contrasts provide the seed gravitational wells into which baryons
fall. Although the relative amounts of nonbaryonic dark matter are reduced
in the straightforward interpretation of the \name\ scenario, there would be
likely to be few changes to the basic qualitative scenario of the initiation of
standard structure formation. If one wishes to completely eliminate
nonbaryonic dark matter on the other hand, then one faces the formidable
challenge of explaining how realistic structures can form from density
contrasts which are only of order $10^{-5}$ at last scattering. Thus the \name\
scenario with the difference, $\OmMn-\OmBn$, interpreted as a nonbaryonic dark
matter component remains the simplest scenario from the viewpoint of present
understanding.

\section{Discussion}

Any serious physical theory should not only be founded on sound principles,
but also provide predictions that can potentially rule it out. Much of the
present review has therefore concentrated on several tests which might
distinguish the \name\ model from models of homogeneous dark energy. The
(in)homogeneity test of Clarkson, Bassett and Lu is a definitive test
independent of the \name\ model with the potential to falsify the standard
cosmology on large scales, since it tests the validity of the Friedmann
equation directly. It would similarly rule out any modified gravity model
which relied on a homogeneous geometry with a Friedmann--type equation
at the largest scales.

In performing any tests, however, one must be very careful to ensure
that data has not been reduced with built--in assumptions that use the
Friedmann equation. For example, current estimates of the BAO scale, such as
that of Percival \etal\ \cite{Percival}, do not determine $D\Z V$
directly from redshift and angular diameter measures, but first
perform a Fourier space transformation to a power spectrum, assuming
a FLRW cosmology. Redoing such an analysis for the \name\ model may
involve a recalibration of relevant transfer functions.

In the case of supernovae, one must also take care since compilations such as
the Union \cite{union}, Constitution \cite{Hicken} and Union2 \cite{union2}
datasets use the SALT or SALT-II methods to calibrate light curves. In this
approach empirical light curve parameters and cosmological parameters -- {\it
assuming the Friedmann equation} -- are simultaneously fit by analytic
marginalisation before the raw apparent magnitudes are recalibrated. As Hicken
\etal\ discuss \cite{Hicken}, a number of systematic discrepancies exist
between data reduced by the different methods even
within the \LCDM\ model. In the case of the \name\ model, we find considerable
differences between the different approaches \cite{SW}, which appear to be
largely due to systematic issues in distinguishing
reddening by host galaxy dust from an intrinsic colour variation in the
supernovae. It is also crucial for the \name\ scenario that data is cut at the
scale of statistical homogeneity ($z\goesas0.033$), below which a simple
average Hubble law is not expected. For datasets reduced by the SALT or SALT-II
methods there is generally Bayesian evidence that favours the \LCDM\ model
over the TS model. By contrast for datasets reduced by MLCS2k2 the
Bayesian evidence favours the TS model over the \LCDM\ model \cite{SW}. In
principle, with perfect standard candles there are already enough supernovae
to decide between the \LCDM\ and \name\ models
on Bayesian evidence, but in practice one is led to different conclusions
depending on how the data is reduced. It is therefore important that the
systematic issues are unravelled.

The value of the dressed Hubble constant is also an observable quantity of
considerable interest. A recent determination of $\Hm$ by Riess \etal\
\cite{shoes} poses a challenge for the \name\ model. However, it is a
feature of the \name\ model that a 17--22\% variance in the apparent Hubble
flow will exist on local scales below the scale of statistical homogeneity,
and this may potentially complicate calibration of the cosmic distance
ladder. Further quantification of the variance in the apparent Hubble
flow in relationship to local cosmic structures would provide an
interesting possibility for tests of the \name\ cosmology for which there
are no counterparts in the standard cosmology.\smallskip

A huge amount of work remains to be done to develop the timescape scenario
to the level of detail of the standard cosmology. At the
mathematical level, we need to refine the notion of coarse--graining of dust
in relation to the various scales of averaging, slicings by hypersurfaces
in the evolution equations, and null cone averages. Whatever the outcome
of such investigations, I find it exciting that much remains still to be
explored in general relativity.

As long as the number of alternative theories is comparable to the number of
``alternative'' theorists, the detailed development of any alternative
paradigm to the standard cosmology may take several years or decades, a
timescale which also applies to the big science projects needed to perform
precision observations such as redshift-time drift test \cite{SML}.
Since one can always achieve better fits by adding new terms to relevant
equations, every theorist is inevitably guided by intuition and aesthetic
judgements about physical principles as much as by existing observations and
experiments.

My own theoretical prejudices are rooted in the knowledge that general
relativity is on one hand an extremely successful theory of nature, in
complete agreement with observations on the scale of stellar systems, and yet
on the other hand, although it is based on deep physical principles, it is
still also a theory which has not been completely understood in terms of the
coarse-graining of dust, averaging, fitting, the statistical nature of
gravitational energy and entropy, and the nature of Mach's principle. Although
the nonlinearities of the Einstein equations may play a role in unravelling the
mystery of dark matter \cite{CT1,CT2}, my own opinion is that what is at stake
is more than simply nonlinear mathematics, but also deep and subtle questions
of physical principle.

Even if the retro-fit of a density distribution to observed galaxy rotation
velocities via Einstein's equations \cite{CT1,CT2} could be independently shown
to closely match the observed density distribution, there may be more subtle
issues relating to coarse--graining and averaging which underlie the formation
of the observed dust distributions, which may also be phenomenologically
applicable to galaxies or galaxy clusters with less symmetry. It is worth
noting that MOG \cite{mog} operates by an effective phenomenological variation
of Newton's constant. Since direct observations never directly involve $G$ but
rather $GM$, my suspicion is that the phenomenology is pointing to the thorny
issue of the definition of gravitational energy when averaging
on different scales. This is the question we need to think more deeply about.
The difficult problem of quasi-local gravitational energy in Einstein's
theory may turn out not to simply be an arcane curiosity in mathematical
relativity, but to be of direct importance for understanding the large
scale structure of the universe.
\section*{Acknowledgments} I thank Prof.\ Remo Ruffini and
ICRANet for support and hospitality while the work of refs.~\cite{equiv,obs}
was undertaken. This work was also partly supported by the Marsden fund
of the Royal Society of New Zealand. I am grateful to many colleagues
and students for numerous discussions, including in particular Thomas Buchert,
Teppo Mattsson, Roy Kerr and Peter Smale. I also thank John Auping for
correspondence which led to the inclusion of the discussion about nonbaryonic
dark matter.
\bibliographystyle{aipproc}

\end{document}